\documentclass[11pt]{article}
\usepackage{amssymb}\usepackage{amsmath}
\usepackage[all]{xy}
\usepackage[dvips]{graphicx}
\numberwithin{equation}{section}

\def\zbar{\bar z}
\def\wbar{\bar w}
\def\xbar{\bar x}
\def\be{\begin{equation}}
\def\ee{\end{equation}}
\def\ba{\begin{align}}
\def\ea{\end{align}}
\def\beq{\begin{eqnarray}}
\def\eeq{\end{eqnarray}}
\def\p{\partial}

\def\yboxit#1#2{\vbox{\hrule height #1 \hbox{\vrule width #1
\vbox{#2}\vrule width #1 }\hrule height #1 }}
\def\fillbox#1{\hbox to #1{\vbox to #1{\vfil}\hfil}}
\def\ybox{{\lower 1.3pt \yboxit{0.4pt}{\fillbox{8pt}}\hskip-0.2pt}}
\def\comments#1{}

\def\p{\partial}

\def\II{\relax{I\kern-.10em I}}

\def\IZ{\relax\ifmmode\mathchoice
{\hbox{\cmss Z\kern-.4em Z}}{\hbox{\cmss Z\kern-.4em Z}}
{\lower.9pt\hbox{\cmsss Z\kern-.4em Z}}
{\lower1.2pt\hbox{\cmsss Z\kern-.4em Z}}\else{\cmss Z\kern-.4em
Z}\fi}
\def\IB{\relax{\rm I\kern-.18em B}}
\def\IC{{\relax\hbox{$\inbar\kern-.3em{\rm C}$}}}
\def\ID{\relax{\rm I\kern-.18em D}}
\def\IE{\relax{\rm I\kern-.18em E}}
\def\IF{\relax{\rm I\kern-.18em F}}
\def\IG{\relax\hbox{$\inbar\kern-.3em{\rm G}$}}
\def\IGa{\relax\hbox{${\rm I}\kern-.18em\Gamma$}}
\def\IH{\relax{\rm I\kern-.18em H}}
\def\II{\relax{\rm I\kern-.18em I}}
\def\IK{\relax{\rm I\kern-.18em K}}
\def\IN{\relax{\rm I\kern-.18em N}}
\def\IP{\relax{\rm I\kern-.18em P}}

\def\inbar{\,\vrule height1.5ex width.4pt depth0pt}

\def\p{\partial}

\font\cmss=cmss10 \font\cmsss=cmss10 at 7pt
\def\IR{\relax{\rm I\kern-.18em R}}

\def\lp10{l_P^{10}}
\def\lp11{l_P^{11}}
\def\R11{R_{11}}

\font\manual=manfnt
\def\dbend{\lower3.5pt\hbox{\manual\char127}}

\def\IZ{\relax\ifmmode\mathchoice
{\hbox{\cmss Z\kern-.4em Z}}{\hbox{\cmss Z\kern-.4em Z}}
{\lower.9pt\hbox{\cmsss Z\kern-.4em Z}} {\lower1.2pt\hbox{\cmsss
Z\kern-.4em Z}}\else{\cmss Z\kern-.4em Z}\fi}

\def\p{\partial}

\def\bar{\overline}

\def\rt2{\sqrt{2}}
\def\irt2{\frac{1}{\sqrt{2}}}

\font\cmss=cmss10
\font\cmsss=cmss10 at 7pt
\def\IL{\relax{\rm I\kern-.18em L}}
\def\IH{\relax{\rm I\kern-.18em H}}
\def\IR{\relax{\rm I\kern-.18em R}}
\def\inbar{\vrule height1.5ex width.4pt depth0pt}
\def\IC{\relax\hbox{$\inbar\kern-.3em{\rm C}$}}
\def\rlx{\relax\leavevmode}
\def\ZZ{\rlx\leavevmode\ifmmode\mathchoice{\hbox{\cmss Z\kern-.4em Z}}
 {\hbox{\cmss Z\kern-.4em Z}}{\lower.9pt\hbox{\cmsss Z\kern-.36em Z}}
 {\lower1.2pt\hbox{\cmsss Z\kern-.36em Z}}\else{\cmss Z\kern-.4em
 Z}\fi}
\def\IZ{\relax\ifmmode\mathchoice
{\hbox{\cmss Z\kern-.4em Z}}{\hbox{\cmss Z\kern-.4em Z}}
{\lower.9pt\hbox{\cmsss Z\kern-.4em Z}}
{\lower1.2pt\hbox{\cmsss Z\kern-.4em Z}}\else{\cmss Z\kern-.4em
Z}\fi}

\font\manual=manfnt
\def\dbend{\lower3.5pt\hbox{\manual\char127}}

\def\IZ{\relax\ifmmode\mathchoice
{\hbox{\cmss Z\kern-.4em Z}}{\hbox{\cmss Z\kern-.4em Z}}
{\lower.9pt\hbox{\cmsss Z\kern-.4em Z}} {\lower1.2pt\hbox{\cmsss
Z\kern-.4em Z}}\else{\cmss Z\kern-.4em Z}\fi}

\def\bar{\overline}

\def\rt2{\sqrt{2}}
\def\irt2{\frac{1}{\sqrt{2}}}

\title{\Large{\bf  Conformal Current Algebra in Two Dimensions}} 
\author{Sujay K. Ashok$^{a,b}$, Raphael Benichou$^{c}$
  and Jan Troost$^{c}$ } \date{}
\begin{document}
\maketitle

\begin{center}
  $^{a}$Institute of Mathematical Sciences\\
  C.I.T Campus, Taramani\\
  Chennai, India 600113\\
  \vspace{.3cm}
  $^{b}$Perimeter Institute for Theoretical Physics\\
  Waterloo, Ontario, ON N$2$L$2$Y$5$, Canada \\
  \vspace{.3cm}
  $^{c}$Laboratoire de Physique Th\'eorique \\
Unit\'e Mixte du CRNS et
    de l'\'Ecole Normale Sup\'erieure \\ associ\'ee \`a l'Universit\'e Pierre et
    Marie Curie 6 \\ UMR
    8549  \footnote{Preprint LPTENS-09/16.} \\ \'Ecole Normale Sup\'erieure \\
  $24$ Rue Lhomond Paris $75005$, France
\end{center}

 \begin{abstract}
   We construct a non-chiral current algebra in two dimensions
   consistent with conformal invariance. We show that the conformal
   current algebra is realized in non-linear sigma-models on
   supergroup manifolds with vanishing Killing form, with or
   without a Wess-Zumino term. The current algebra is computed using
   two distinct methods. First we exploit special algebraic properties of
   supergroups to compute the exact two- and three-point functions of
   the currents and from them we infer the current algebra. The 
   algebra is also calculated by using conformal perturbation
   theory about the Wess-Zumino-Witten point and resumming the
   perturbation series. We also prove that these models realize a
   non-chiral Kac-Moody algebra and construct an infinite set of
   commuting operators that is closed under the action of the
   Kac-Moody generators. The supergroup models that we consider
   include models with applications to statistical mechanics,
   condensed matter and string theory. In particular, our results may
   help to systematically solve and clarify the quantum integrability
   of $PSU(n|n)$ models and their cosets, which appear prominently in
   string worldsheet models on anti-deSitter spaces.
\end{abstract}

\newpage

\tableofcontents

\section{Introduction}
Two-dimensional sigma-models on supergroups have applications to a
wide range of topics such as the integer quantum hall effect, quenched
disorder systems, polymers, string theory, as well as other domains in
physics (see e.g. \cite{PS, Efetov:1983xg, Sethi:1994ch,
  Zirnbauer:1999ua, Guruswamy:1999hi}). The principal chiral model is
perturbatively conformal on various supergroups
\cite{Bershadsky:1999hk, Babichenko:2006uc} with or without the
addition of a Wess-Zumino term, and at least to two loop order on
their cosets with respect to a maximal regular subalgebra
\cite{Babichenko:2006uc}.
Sigma models on graded supercosets are also believed to be conformal \cite{Kagan:2005wt}.
 Thus, these models have an infinite
dimensional symmetry algebra that should tie in with their supergroup
symmetry. An extended non-linear symmetry algebra was identified in
\cite{Bershadsky:1999hk} but the representation theory of the algebra
seems difficult to establish. Steps towards solving these models were
made using various techniques \cite{Read:2001pz, Gotz:2006qp,
  Quella:2007sg}. In this paper, we exhibit a  conformal current
algebra in these models. The algebra of currents is non-chiral and implies 
conformal symmetry, hence the name.

The models under discussion enter as the key building blocks in
worldsheet sigma-models on supersymmetric $AdS$ backgrounds in string
theory.  The supergroup $PSU(1,1|2)$ principal chiral model
corresponds to a supersymmetric $AdS_3 \times S^3$ background with
Ramond-Ramond flux \cite{Bershadsky:1999hk, Berkovits:1999im}.  Since a theory of quantum gravity on asymptotically $AdS_3$ space-times, supplemented with appropriate
boundary conditions, exhibits an infinite dimensional conformal
symmetry algebra \cite{Brown:1986nw}, we should be able to construct
those generators from the worldsheet theory.  Indeed, for $AdS_3$
string theory with only Neveu-Schwarz-Neveu-Schwarz flux, it has been
shown how to construct the space-time Virasoro algebra in terms of the
worldsheet current algebra \cite{Giveon:1998ns, Kutasov:1999xu, de
  Boer:1998pp}. To perform a similar construction in Ramond-Ramond
backgrounds, one needs to understand the worldsheet current algebra
for two-dimensional models with supergroup targets.

A second application within this context is the extension of our
analyis to supercoset manifolds, which includes the $AdS_5 \times S^5$
background of string theory. The worldsheet current algebra is tied in
with the integrability of the worldsheet theory \cite{Bena:2003wd}. Our work 
may help in systematically exploiting the integrability of
the worldsheet model at the quantum level, with applications to the
solution of the spectrum of planar four-dimensional gauge theories (see e.g.
\cite{Gromov:2009tv}) via the $AdS/CFT$ correspondence.

The plan of the paper is as follows.  In section \ref{generic} we kick
off with a general analysis of two-dimensional current algebra
operator product expansions that are consistent with locality, Lorentz
invariance and parity-time reversal.  We exhibit a particular current
algebra that is non-chiral and consistent with conformal invariance.
We then move on to exhibit a realization of the algebra in a 
conformally invariant  model
on a supergroup manifold.

We calculate perturbatively the current two- and three-point functions of
these models in section \ref{correlators}. From these correlators we
infer the operator product expansions of the currents, which are shown
to fall into the conformal current algebra class discussed in section
\ref{generic}. In section \ref{firstorder} we analyze deformed
Wess-Zumino-Witten models on supergroups using conformal perturbation
theory. We compute the operator product expansions of the currents to
all orders and resum the series, thereby demonstrating that the
resulting algebra matches the results obtained using purely algebraic
properties of the supergroups.  We analyze the current algebra on the
cylinder in terms of a Fourier decomposition in section \ref{cylinder}
and show the existence of a Kac-Moody subalgebra and an infinite
set of commuting operators that transform amongst each other under
the Kac-Moody subalgebra. In section
\ref{conclusions} we discuss some applications of the conformal
current algebra that we have found and possible future directions of
work.

\section{Current algebras in two dimensions}
\label{generic}
Before we compute the current algebra operator product expansion
\cite{Wilson:1969zs} for supergroup sigma-models, it is interesting to
analyze the generic operator product expansions (OPEs) involving the
currents for a two-dimensional model which is local, Lorentz invariant
and which respects parity-time reversal.

Previously, the generic two-dimensional current algebra was analyzed in
\cite{Luscher:1977uq} where it was applied to an asymptotically free
sigma-model -- the $O(n)$ sigma-model. Parity symmetry was assumed to
be valid in the analysis. Later, a similar analysis was
performed \cite{Bernard:1990jw, Bernard:1991zy} for massive
models with Wess-Zumino-Witten ultraviolet fixed points. In both cases, the
study was applied to argue for the integrability of the two-dimensional
sigma-model in the quantum theory.

The models that we will study have the distinctive feature that they are
conformal. Moreover, the Wess-Zumino term breaks parity invariance.
Therefore, we start by analyzing the generic current operator product
expansions consistent with locality, Lorentz invariance and
PT-invariance only. In the following, we generalize the
methodology of \cite{Luscher:1977uq}.

\subsection{Locality, Lorentz invariance and PT-invariance}

In the absence of parity invariance, the vector representation of the
two-dimensional Lorentz group splits into two one-dimensional
irreducible representations. A current $j_\mu$ can therefore be split
into two irreducible representations $j_+$ and $j_-$ of the
two-dimensional Lorentz group\footnote{We work in Lorentzian signature
  in this section. We define $x^{\pm} = x \pm t$ and $ds^2=-dt^2+dx^2
  =dx^+ dx^-$. We denote $x^2=x^+ x^-$.}. We write the OPEs in terms
of these irreducible components, leading to three independent OPEs,
between the pairs of current components $(j_+, j_+$), $(j_+, j_-)$,
and $(j_-, j_-)$.

We first analyze the OPE between the components 
$j_+$ and $j_+$. We take the currents
to be in the adjoint representation of a symmetry group $G$ of the
model: $j_+ = j_+^a t_a$ where $t_a$ spans the Lie algebra of $G$.
We write down the generic OPE in terms of Lorentz invariant coefficient 
functions. Moreover, we assume that the only low-dimensional
 operators that appear in the operator
product expansion are the identity operator, the currents and their
derivatives.
We also assume that the currents have conformal dimension one. Thus the previous list of allowed operators in the OPE should account for all the terms up to regular ones.
The $j_+ j_+$ OPE is then given by:
\begin{multline}
j_+^a (x) j_+^b (0)  \sim  \alpha^{ab} 
(x^-)^2 d_1(x^2)  + {\beta^{ab}}_c (x^-)^2 \left[ d_2(x^2) x^+ j_+^c(0)+d_3(x^2) x^- j_-^c(0) \right.
\cr
 \left. + e_1(x^2) x^+ x^-\partial_+ j_-^c(0) + e_2(x^2) x^+ x^- \partial_- j_+^c(0) 
\right. \cr \left. 
+ e_3(x^2) (x^+)^2 \partial_+ j_+^c(0)
+ e_4(x^2) (x^-)^2 \partial_- j_-^c(0) \right] 
  + \dots
\end{multline}
where the functions $d_i,e_i$ are functions of the Lorentz
invariant $x^2 = x^+ x^-$. The tensor $\alpha^{ab}$ is an invariant
two-tensor in the product of adjoint representations, and
${\beta^{ab}}_c$ represents an adjoint representation in the
product of two adjoints.  We will assume that $\alpha^{ab}$
corresponds to 
a non-degenerate bi-invariant 
metric $\kappa^{ab}$ and that the tensor
${\beta^{ab}}_c$ is equal to the structure constants ${f^{ab}}_c$ of
the Lie algebra of $G$. 
 We take the structure constants to be
anti-symmetric in its indices\footnote{More precisely, we take the
  structure constants to be graded anti-symmetric when
$G$ is a supergroup. Although the
  grading is crucial, it will not affect our formulas, except for a
  plethora of minus signs when interchanging operators -- these will not
  influence our final results much. We maintain consistency
with the grading throughout section 2, but not necessarily through the rest of
the paper. }. We use the fact that we can
interchange operators\footnote{Up to a minus sign for fermionic
  operators. We see, for example, that the interchange of fermionic operators
will cancel a minus sign from the grading of the superalgebra when $G$ is 
a supergroup. } to determine that the above OPE should be equal to:
\begin{align}
j_+^b(0) j_+^a(x) &  \sim  \kappa^{ba} 
(x^-)^2 d_1  - {f^{ba}}_c (x^-)^2 (d_2 x^+ j_+^c(x)+d_3 x^- j_-^c(x)
\cr
& - e_1 x^+ x^- \partial_+ j_-^c(x) - e_2 x^+ x^- \partial_- j_+^c(x) - e_3 (x^+)^2 \partial_+ j_+^c(x)
- e_4 (x^-)^2 \partial_- j_-^c(x))
\cr
&  \sim  \kappa^{ba} 
x^- x^- d_1    - {f^{ba}}_c x^- x^- (d_2 x^+ j_+^c(0)+d_3 x^- j_-^c(0)
\cr 
& \hspace{.5in} +(d_3- e_1) x^+ x^- \partial_+ j_-^c(x) +(d_2- e_2) x^+ x^- \partial_- j_+^c(x)
\cr
&\hspace{1in}  +(d_2- e_3) x^+ x^+ \partial_+ j_+^c(x) +(d_3- e_4) x^- x^- \partial_- j_-^c(x))
\end{align}
from which we derive the equations:
\begin{eqnarray}
e_1 = d_3/2  & \qquad &
e_2 = d_2/2 \nonumber \\
e_3 = d_2/2 & \qquad &
e_4 = d_3/2,
\end{eqnarray}
which gives rise to the simplified operator product expansion:
 \begin{align}
 j_+^a (x) j_+^b (0) & \sim  \kappa^{ab} 
 (x^-)^2 d_1 + {f^{ab}}_c (x^-)^2 \left[ d_2 x^+ j_+^c(0)+d_3 x^- j_-^c(0) + \frac{d_3}{2} x^+ x^-\partial_+ j_-^c(0) \right.
 \cr
 &  \left. + \frac{d_2}{2} x^+ x^- \partial_- j_+^c(0) +\frac{d_2}{2}  (x^+)^2 \partial_+ j_+^c(0)
 +\frac{d_3}{2} (x^-)^2 \partial_- j_-^c(0) \right]  + 
\dots
 \end{align}
The OPE has one free coefficient $d_1$ (function of $x^2$)
 at leading order, and two at subleading order.
We similarly obtain:
 \begin{align}
 j_-^a (x) j_-^b (0) & \sim  \kappa^{ab} 
 (x^+)^2 d_4 + {f^{ab}}_c (x^+)^2  \left[ d_5 x^- j_-^c(0) + d_6 x^+ j_+^c(0)+ \frac{d_5}{2} x^+ x^-\partial_+ j_-^c(0) \right.
\cr
& \left. +\frac{d_5}{2} (x^-)^2 \partial_- j_-^c(0) + \frac{d_6}{2} x^+ x^- \partial_- j_+^c(0) +\frac{d_6}{2}  (x^+)^2 \partial_+ j_+^c(0)
  \right]  + \dots
 \end{align}
For the OPE between $j_+$ and $j_-$ we don't get as many constraints. We find $7$ more free functions:
\begin{align}
j_+^a (x) j_-^b(0) & \sim  \kappa^{ab} 
d_7 + {f^{ab}}_c  \left[ d_8 x^+ j_+^c(0)+d_9 x^- j_-^c(0) 
 + d_{13} x^+ x^-\partial_+ j_-^c(0) + d_{12} x^+ x^- \partial_- j_+^c(0) \right.  \cr 
&\left. \hspace{1in}+d_{10}  (x^+)^2 \partial_+ j_+^c(0)
+d_{11} (x^-)^2 \partial_- j_-^c(0) \right]  + \dots
\end{align}
We have a total of $13$ free coefficient functions.

\subsection{Current conservation}
We now impose consistency of the operator product expansions of the
currents with current conservation.  We choose the relative normalization of
the two components of the currents such that the equation of current
conservation reads \footnote{We expect current
  conservation to only hold up to delta-function contact terms. 
We therefore do not keep track
  of contact terms in the following. }:
\begin{eqnarray}\label{consCurrent}
\partial^\mu j_\mu^a = \partial_- j_+^a + \partial_+ j_-^a &=& 0.
\end{eqnarray}
Current conservation implies that one of the coefficients in the $j_+
j_-$ OPE (namely $d_{12}+d_{13}$) becomes redundant. We check that the
OPEs of the current conservation equation (\ref{consCurrent}) with the
currents $j_+ $ and $j_-$ vanish.  That leads to the set of equations
\footnote{There is no equation corresponding to the operator
  $\partial_\mu j^\mu$ since the coefficient multiplies zero.}:
\begin{align}\label{EOMcheck}
& d_1 + \frac{x^2}{2} d_1' +\frac{1}{2} d_7' = 0 \,, \qquad
d_2 +  \frac{x^2}{2} d_2' +\frac{1}{2} d_8' + \frac{d_8}{2 x^2} = 0\,, \cr
& d_4 + \frac{x^2}{2} d_4' + \frac{1}{2} d_7' = 0\,, \qquad
\frac{3}{2} d_6 + \frac{x^2}{2} d_6' +\frac{1}{2} d_8' = 0\,, \cr
&\frac{3}{2} d_3 +  \frac{x^2}{2} d_3' +\frac{1}{2} d_9' = 0\,, \quad 
2 d_3 +  \frac{x^2}{2} d_3' +  (d_9'-d_{11}') = 0\, \cr
&2 d_6 + \frac{x^2}{2} d_6' +  d_{10}' = 0\,, \qquad 
d_5 + \frac{x^2}{2} d_5' +  d_{11}' + \frac{2}{x^2} d_{11} = 0\,,\cr
&d_5 + \frac{x^2}{2} d_5' +\frac{1}{2} d_9' + \frac{1}{2 x^2} d_9 = 0\,, \cr
& d_2 +  \frac{x^2}{2} d_2' + (d_8'-d_{10}') + \frac{2}{x^2} (d_8-d_{10}) = 0\cr
&\frac{3}{2} (d_6-d_5) + \frac{x^2}{2} (d_6'-d_5') + (d_{12}'-d_{13}')+ \frac{1}{x^2} (d_{12}-d_{13}) = 0\,,\cr
& \frac{3}{2}(d_2-d_3)+  \frac{x^2}{2} (d_2'-d_3') + (d_8'-d_{12}'-(d_9'-d_{13}'))+ \frac{1}{x^2} (d_8-d_{12}-(d_9-d_{13})) = 0\,.
\end{align}
We get a system of 
twelve first-order differential equations for
twelve functions. We will not try to solve them in full generality.
Instead, we assume that the leading and  subleading
singularities in the OPEs are powerlike. This leads to the following
ansatz:
\begin{align}
&d_1(x^2) = c_1 /x^4, \quad 
d_2(x^2) = c_2 /x^4,  \quad
d_3(x^2) = f_1 /x^4\,, \cr 
&d_4(x^2) = c_3 /x^4, \quad
d_5(x^2) = c_4 /x^4,\quad
d_6(x^2) = f_7 /x^4, \cr
&d_7(x^2) = f_2 /x^2,\quad
d_8(x^2) = f_3 /x^2,\quad
d_9(x^2) = f_4 /x^2\, .
\end{align}
The coefficient $c_i$'s and $f_i$'s are now constant coefficients.
Plugging this ansatz into the equations \eqref{EOMcheck}, we get the
following relations between the coefficients:
\begin{align}
f_2 = 0, \qquad 
f_4 = f_1, \qquad 
f_3 = f_7 = c_4 - c_2 + f_1\,.
\end{align}
The remaining equations in \eqref{EOMcheck} then allow to solve for
the subsubleading coefficient functions in the OPEs:
\begin{align}\label{diffd12d13}
&d_{10}(x^2) = \frac{f_7}{x^2}, \qquad  
d_{11}(x^2) = 0, \cr
& (d_{12}'-d_{13}')
+ \frac{1}{x^2} (d_{12}-d_{13}) =  \frac{c_2-f_1}{2x^4}.
\end{align}
So it only remains to solve for $d_{12}(x^2) - d_{13}(x^2)$ in terms of the $c_i$ and $f_1$. 
Denoting $c_2-f_1=g$, the general solution to the differential
equation \eqref{diffd12d13} reads: \be d_{12}(x^2)-d_{13}(x^2) =
\frac{2 c_5}{x^2}+\frac{g}{2x^2} \log \mu^2 x^2 \ee
where $c_5$ is a constant coefficient and $\mu$ is an arbitrary mass
scale. We note that when the coefficient $g$ is non-zero, we can
absorb the coefficient $c_5$ in a redefinition of the mass scale
$\mu$. So the OPEs are given in terms of five dimensionless
coefficients. They read:
\begin{align}
j_+^a (x) j_+^b (0) & \sim  \frac{\kappa^{ab} c_1 }{(x^+)^2}  + {f^{ab}}_c \left[ \frac{c_2}{x^+} j_+^c(0)+  \frac{(c_2-g)x^-}{(x^+)^2} j_-^c(0) - \frac{g}{4} \frac{x^-}{x^+} \left( \partial_+ j_-^c(0) - \partial_- j_+^c(0) \right) \right. \cr
&\hspace{1.5in} \left. 
 +\frac{c_2}{2} \partial_+ j_+^c(0)
+\frac{c_2-g}{2} \frac{(x^- )^2}{(x^+)^2} \partial_- j_-^c(0)\right] 
+ \dots
\cr
j_-^a (x) j_-^b (0) & \sim   \frac{\kappa^{ab} 
c_3}{(x^-)^2} + {f^{ab}}_c \left[ \frac{c_4}{x^-} j_-^c(0) + \frac{(c_4-g)\, x^+}{(x^-)^2} j_+^c(0)+  \frac{g}{4} \frac{x^+}{x^-} \left( \partial_+ j_-^c(0) - \partial_- j_+^c(0) \right)\right. \cr
&\left. \hspace{1.5in} +\frac{c_4}{2} \partial_- j_-^c(0)
+\frac{c_4-g}{2} \frac{(x^+)^2}{(x^-)^2} \partial_+ j_+^c(0) \right] + \dots
\cr
j_+^a (x) j_-^b(0) & \sim   {f^{ab}}_c  \left[ \frac{c_4-g}{x^-} j_+^c(0)
+ \frac{(c_2-g)}{x^+} j_-^c(0) + \frac{(c_4-g)x^+}{x^-} \partial_+ j_+^c (0) \right.
\cr
&\left. \hspace{1.1 in}- \left( c_5 +\frac{g}{4}  \log \mu^2 x^2 \right) \left( \partial_+ j_-^c(0) 
- \partial_- j_+^c(0)\right) \right] 
+ \dots
\end{align}
It would be interesting to search for more general solutions to the set of differential equations.

\subsection{The Maurer-Cartan equation}
In this subsection we show that under certain circumstances, we can
obtain further constraints on the current algebra.  Consider a field
$g$ taking values in a Lie group. The one-form $dg g^{-1}$ satisfies
the Maurer-Cartan equation \be d(dg g^{-1}) = dg g^{-1} \wedge dg
g^{-1}. \ee We will get further constraints  if 
we suppose that the components of the current are
related to the field $g$ in the following way: \be j_+ =c_+ \partial_+
g g^{-1}, \quad j_- =c_- \partial_- g g^{-1} \ee where $c_+$ and $c_-$
are constant coefficients.
 The generators of the Lie 
(super-)%
algebra satisfy: ${[} t_a , t_b {]}= i {f^c}_{ab} t_c$.
Then the Maurer-Cartan
equation takes the form
\begin{eqnarray}
c_- \partial_- j^a_+ 
- c_+ \partial_+ j_-^a - i  {f^a}_{bc}  j_+^c j_-^b  &=& 0.
\end{eqnarray}
We want to ensure that this equation is also valid in the quantum
theory. However the operator defined as the product of two currents
needs to be regularized. For this reason the validity of the
Maurer-Cartan equation in the quantum theory requires more discussion.

\subsubsection*{Normal Ordering}
In the quantum theory, we introduce a normal ordering for composite
operators based on a point-splitting procedure.  The normal ordered
product $:O_1 O_2 :(y) $ of two operators $O_1$ and $O_2$ evaluated
at the point $y$ is defined to be the product of the operator $O_1$ at
the point $x$ with the operator $O_2$ at the point $y$, in the limit
as $x$ approaches $y$. The regularization amounts to dropping the terms that are singular in this limit. For this procedure to be well-defined, it is important that the resulting operator is evaluated at the point $y$.  We will denote this procedure by
\be
:O_1 O_2:(y) = \lim_{:x\rightarrow y:}O_1(x) O_2(y) \,.
\ee
We note that the operators within the normal 
ordered product $:O_1 O_2:$ do not commute\footnote{See e.g. \cite{yellowbook} for a discussion of this fact in the context of chiral conformal field theories.}.
We will later confirm that a natural choice for the normal ordered Maurer-Cartan equation in this scheme is\footnote{We use the notation $(-)^a=+1$ if $a$ is a bosonic index, and $-1$ if $a$ is a fermionic index.}:
\begin{eqnarray}\label{:MC:}
c_- \partial_- j^a
_+ 
- c_+ \partial_+ j_-^a - \frac{i}{2} {{f^a}_{bc}}  (:j_+^c j_-^b:+ (-)^{bc} :j_-^b j_+^c:)  &=& 0.
\end{eqnarray}

\subsubsection*{Additional constraints from the Maurer-Cartan equation}
As for the current conservation equation, we ask for the OPE between
the quantum Maurer-Cartan equation \eqref{:MC:} and the current to
vanish.  The first non-trivial constraint is obtained for the
subleading terms.
This leads to 
a relation between the
coefficient of the current algebra $c_1$, $c_2$ and $g$, and the
coefficients $c_+$ and $c_-$: \be\label{MCone}
(c_+ + c_-)(c_2-g)+i c_1= 0 \,. \ee 
We similarly find a constraint linking $c_3$, $c_{4}$ and $g$ to $c_+$
and $c_-$: 
\be\label{MCtwo}
(c_++c_-)(c_4-g)+ic_3  =0 \,.  
\ee
When we consider concrete models realizing the current
algebra, the coefficients $c_+$ and $c_-$ can be derived from the
Lagrangian. The Maurer-Cartan equation then reduces
the number of free constant coefficients from five to three.

\subsection{The Euclidean  current algebra}

For future purposes, we wish to translate the result we obtained for
the current operator algebra into euclidean signature. We perform the
Wick rotation $t \rightarrow -i \tau$, and define the complex
coordinates $z = x- i \tau$, $\bar{z} = x+i \tau$. The current algebra
OPEs become
\begin{align}\label{euclidOPEs}
  j_z^a (z) j_z^b (0) & \sim \kappa^{ab} \frac{c_1}{z^2} + {f^{ab}}_c
  \left[ \frac{c_2}{z} j_z^c(0)+ (c_2-g) \frac{\bar{z}}{z^2}
    j_{\bar{z}}^c(0) \right.\cr & \hspace{1cm} \left. - \frac{g}{4}
    \frac{\bar{z}}{z} (\partial_z j_{\bar{z}}^c(0)
    - \partial_{\bar{z}} j_z^c(0)) +\frac{c_2}{2} \partial_z j_z^c(0)
    +\frac{c_2-g}{2} \frac{\bar{z}^2}{z^2} \partial_{\bar{z}}
    j_{\bar{z}}^c(0) \right] + \dots
  \cr j_{\bar{z}}^a (z) j_{\bar{z}}^b (0) & \sim \kappa^{ab} c_3
  \frac{1}{\bar{z}^2} + {f^{ab}}_c \left[ \frac{c_4}{\bar{z}}
    j_{\bar{z}}^c(0) + \frac{(c_4-g)z}{\bar{z}^2} j_z^c(0) \right. \cr
  & \hspace{1cm} \left. + \frac{g}{4} \frac{z}{\bar{z}} (\partial_z
    j_{\bar{z}}^c(0) - \partial_{\bar{z}} j_z^c(0))
    +\frac{c_4}{2} \partial_{\bar{z}}
    j_{\bar{z}}^c(0)+\frac{(c_4-g)}{2}
    \frac{z^2}{\bar{z}^2} \partial_z j_z^c(0)\right] + \dots
  \cr j_z^a (z) j_{\bar{z}}^b(0) & \sim {f^{ab}}_c \left[
    \frac{(c_4-g)}{\bar{z}} j_z^c(0) +\frac{(c_2-g) }{z}
    j_{\bar{z}}^c(0) + \frac{(c_4-g)z}{\bar{z}} \partial_z j_z^c(0)
  \right.\cr &\hspace{1cm} \left. - (c_5 +\frac{g}{4} \log \mu^2 |z|^2)
    (\partial_z j_{\bar{z}}^c (0) - \partial_{\bar{z}} j_z^c(0))
  \right] + \dots
\end{align}
Later on, we will compare the current algebra operator product
expansions in equations (\ref{euclidOPEs}) to those of a supergroup
non-linear sigma-model with Wess-Zumino term. We will find specific
expressions for the coefficients $c_i$ and $g$ in terms of the
parameters in the Lagrangian.

\subsection{Conformal current algebra}

It turns out that the above current algebra can become the building
unit for a conformal algebra when the Killing form of the (super-)group
vanishes\footnote{For simple super Lie algebras the vanishing of the 
Killing form is equivalent to the vanishing of the dual Coxeter number.}. 
In that special case the current algebra is promoted to a
conformal current algebra, namely, the Sugawara stress-energy tensor
built from the currents satisfies the canonical conformal operator
product expansion. 
The terms that in other circumstances
 spoil conformality are eliminated through the fact
that the Killing form is zero.
The holomorphic component of the stress-energy tensor is
\be T(w) = \frac{1}{2 c_1} :j_{bz} j^{b}_{z}:(w), \ee
as we will demonstrate\footnote{ We defined $j_a = j^b \kappa_{ba}$
and $\kappa^{ac} \kappa_{cb}={\delta^a}_b$. In the following we also use
the convention ${f_{abc}}={f_{ab}}^d \kappa_{dc}$.}.

\subsection*{The current as a conformal primary}
First we compute the OPE between the current $j_z^a$ and
 its bilinear combination $: j_{bz} j^b_z :$, using a point-splitting procedure:
\begin{align} 
  j_z^a(z) : j_{bz}j^b_z:(w) &=
  \lim_{:x\to w:} j_z^a(z) \left[ j_{bz}(x) j^{b}_{z}(w) \right] \\
  &= \lim_{:x\to w:} \left[ \left( \frac{c_1 \delta^a_b}{(z-x)^2} +
      {f^{a}}_{bc} \left( \frac{c_2}{z-x} j_z^c(x) +
        \frac{(c_2-g)(\bar z - \bar x)}{(z-x)^2}j_{\bar z}^c(x)
      \right) \right) j^b_{z}(w) \right. \cr & \quad \left. 
+ j_z^d(x) (-1)^{ab} \kappa_{db}
    \left( \frac{c_1 \kappa^{ab}}{(z-w)^2} + {f^{ab}}_{c} \left(
        \frac{c_2}{z-w} j_z^c(w) + \frac{(c_2-g)(\bar z - \bar
          w)}{(z-w)^2}j_{\bar z}^c(w) \right)
    \right)\right]\nonumber\,.
\end{align}
At this point we have to perform the OPE between the operators
evaluated at the point $x$ and the operators evaluated at the point
$w$. Then we take the limit where $x$ goes to $w$ and discard the
singular terms. We notice already that only the regular terms in the
OPEs of the second line will contribute to the final result.  We get
\begin{align} 
  j_z^a(z) :j_{bz} j^{b}_{z}:(w) &= \lim_{:x\to w:} \left[ c_1
    \frac{j_z^a(w)}{(z-x)^2} + \frac{c_2 {f^{a}}_{bc}}{z-x} \left(
      \frac{c_1 \kappa^{cb}}{(x-w)^2} \right. \right. \cr 
& \qquad
  \left. \left.  + {f^{cb}}_d \left(\frac{c_2 j_z^d(w)}{x-w}+
        (c_2-g)\frac{\bar x - \bar w}{(x-w)^2}j_{\bar z}^d(w) \right)
      + :j^c_z j^b_z:(w) \right) \right. \cr 
& \quad \left.  +
    \frac{(c_2-g) {f^{a}}_{bc} (\bar z - \bar x)}{(z-x)^2} \left(
      {f^{cb}}_d \left( \frac{(c_4-g) j_z^d(w)}{\bar x - \bar w} +
        \frac{(c_2-g) j_{\bar z}^d(x)}{x-w}
      \right. \right. \right. \cr 
& \qquad \left. \left. \left.  -
        \left( c_5 +\frac{g}{4} \log \mu^2 |x-w|^2 \right) (\partial_z
        j_{\bar z}^d(w) - \partial_{\bar z} j_z^d(w)) \right) +
      :j_{\bar z}^c j_z^b:(w) \right) \right]\cr 
& + c_1
  \frac{j_z^a(w)}{(z-w)^2} + (-1)^{bc} {f^{a}}_{bc}\frac{c_2}{z-w} :j^b_z
  j^c_z:(w)  
\cr
&
+ (-1)^{bc} {f^{a}}_{bc}\frac{(c_2-g)(\bar z - \bar w)}{(z-w)^2}
  :j^b_z j^c_{\bar z}:(w) \,.\cr
\end{align}
Among the remaining terms, many cancel: every contraction of the 
invariant 
metric with a structure constant gives zero by symmetry, and
the double contractions of two structure constants are proportional to
the Killing form and thus also vanish.
We are left with \beq j_z^a(z) :j_{bz} j^b_z:(w) &=& 2c_1
\frac{j_z^a(w)}{(z-w)^2} + c_2 \frac{ {f^a}_{bc}}{z-w}\left( (-1)^{bc} :j^b_z
  j^c_z:(w)+ :j^c_z j^b_z:(w)\right) \cr && \quad+ (c_2-g) {f^a}_{bc}
\frac{\bar z - \bar w}{(z-w)^2}\left( (-1)^{bc}:j^b_z j^c_{\bar z}:(w)+:
  j^c_{\bar z} j^b_z:(w) \right). \eeq
 The second term vanishes because
of the 
anti-(super)symmetry
of the structure constants. We can simplify the
third term using the Maurer-Cartan identity: \be j_z^a(z) :j_{bz}
j^b_z:(w) = 2c_1 \frac{j_z^a(w)}{(z-w)^2} + 2i (c_2-g) \frac{\bar z -
  \bar w}{(z-w)^2} \left( c_- \bar \partial j_z^a(w) - c_+ \partial
  j_{\bar z}^a(w) \right). \ee By current conservation this can be
rewritten as:

\be j_z^a(z) :j_{bz} j^b_z:(w) = 2c_1 \frac{j_z^a(w)}{(z-w)^2} + 2i
(c_2-g) \frac{\bar z - \bar w}{(z-w)^2} (c_- + c_+) \bar \partial
j_z^a(w). \ee We can now show that the current $j_z^a$ is a primary
field of conformal weight one.  We deduce from the previous computation the
OPE between the stress-energy tensor and the current $j_z^a$, by
expanding the operators on the right-hand side in the neigbourhood of
the point $z$: 
\be 2c_1 T(w) j_z^a(z) = 2c_1
\frac{j_z^a(z)}{(w-z)^2} + 2c_1 \frac{\partial j_z^a(z)}{w-z} +
2(-c_1+i(c_2-g) (c_- + c_+))\frac{\bar z - \bar w}{(z-w)^2}\bar
\partial j_z^a(z) \ee 
Using the relation obtained in equation 
\eqref{MCone}, we finally have
\be 
T(w) j_z^a(z) = \frac{j_z^a(z)}{(w-z)^2} + \frac{\partial j_z^a(z)}{w-z}\,, 
\ee 
which shows that the current $j_z$ is a primary field of conformal dimension one.
It can similarly be checked that $j_{\bar{z}}$ is a conformal primary of dimension zero.

\subsection*{The stress-energy tensor}
We now want to compute the OPE between $T(z) $ and $T(w)$.
This calculation relies on the preceeding calculation
and on the double pole in
the current-current operator product expansion. We get: 
\beq T(z) T(w) &=& \frac{1}{2c_1}
\lim_{:x \to w:} T(z) \left[j_{za}(x) j_z^a(w) \right] \cr &=&
\frac{1}{2c_1} \lim_{:x \to w:} \left[ \left(
    \frac{j_{za}(x)}{(z-x)^2}+ \frac{\partial j_{za}(x)}{z-x}
  \right)j_z^a(w) \right. \cr &&\left. \quad + j_{za}(x)\left(
    \frac{j_{z}^a(w)}{(z-w)^2}+ \frac{\partial j_z^a(w)}{z-w} \right)
\right] \eeq
%
In the second line, only the regular terms in the remaining OPE's will contribute to the final result.
In the first line, all the terms proportional to the structure
constants disappear once again.  We get:
\begin{multline} 
T(z) T(w) =
\frac{1}{2c_1} \lim_{:x \to w:} \left[
\left(
\frac{c_1 \kappa_{ba} \kappa^{ba}}{(z-x)^2(x-w)^2} + \frac{:j_{za} j^a_{z}:(w)}{(z-x)^2} - \frac{2c_1 \kappa_{ba} \kappa^{ba}}{(z-x)(x-w)^3} 
\right. \right. \cr
\left. \left. \quad + \frac{:(\partial j_{za}) j^{a}_z:(w)}{z-x}\right)+ \frac{:j_{za} j^a_{z}:(w)}{(z-w)^2} + \frac{: j_{za} (\partial j^{a}_z):(w)}{z-w} \right] 
\end{multline}
To take the limit, we expand all the functions of $x$ in the
neighbourhood of the point $w$ and keep only the regular term: \be
T(z) T(w) = \frac{dim\ G}{2(z-w)^4} + \frac{2T(w)}{(z-w)^2} +
\frac{\partial T(w)}{z-w}\,, \ee which proves that we have indeed a
conformal algebra of central charge $c= dim\ G$. For supergroups, the
relevant dimension is the superdimension (which is the
self-contraction of the invariant metric).

\subsection*{Summary}
We have shown that a fairly generic current algebra leads to a
conformal theory when the Killing form is zero. The corresponding
stress-energy tensor is given by the Sugawara construction. The
holomorphic current component is a conformal primary with respect to
the holomorphic Virasoro algebra.  The Sugawara energy-momentum tensor
has a central charge equal to the superdimension of the supergroup.

We note that a supergroup with 
zero Coxeter number 
shares some features with
a free theory. The central charge takes its naive value. The composite part of the Maurer-Cartan equation does not need to be renormalized. We will see other simplifications for these models further on.

\section{Current algebra from supergroup current correlators}
\label{correlators}

We now switch gears and consider a concrete model in which the 
generic analysis of two-dimensional current algebras of section
\ref{generic}
can be applied. We consider a conformal supergroup sigma-model
from the list given in  \cite{Babichenko:2006uc}.
 Though we believe our analysis
applies to the whole list, some facts that we use below have been proven explicitly
only for the $PSL(n|n)$ models. We will calculate two-, three- and four-point functions
of currents. Later we will infer the operator algebra of the currents from those correlation functions. 

\subsection{The model}
We consider a supergroup non-linear sigma-model with standard kinetic
term based on a bi-invariant metric on the supergroup. It is the
principal chiral model on the supergroup. In addition we allow for a
Wess-Zumino term. Therefore, we have two coupling constants, namely
the coefficient of the kinetic term $1/f^2$ and the coupling
constant $k$ preceding the Wess-Zumino term. The action is
\footnote{Our normalizations and conventions are mostly as in
  \cite{yellowbook}. In particular we define the primed trace as $Tr'
  (t_a t_b) = 2 \kappa_{ab}$ where $\kappa_{ab}$ is normalized to be
  the Kronecker delta-function for a compact subgroup.  The action is
  written in terms of real euclidean coordinates. We soon switch to
  complex coordinates via $z=x^1+ix^2$. See \cite{yellowbook} for further
  details.  Starting in this section, we will no longer be careful in 
keeping track of the signs
  due to the bosonic or fermionic nature of the super Lie algebra
  generators. They can consistently be restored.}:
\begin{align}
S &= S_{kin} + S_{WZ}\cr
S_{kin} &=  \frac{1}{ 16 \pi f^2}\int d^2 x Tr'[- \partial^\mu g^{-1}
\partial_\mu g]
\cr
S_{WZ} &= - \frac{ik}{24 \pi} \int_B d^3 y \epsilon^{\alpha \beta \gamma}
Tr' (g^{-1} \partial_\alpha g g^{-1} \partial_\beta g   g^{-1} \partial_\gamma g )
\end{align}
Using complex coordinates, and after taking the trace, the kinetic term becomes:
\be
S_{kin} = -\frac{1}{4 \pi f^2} \int d^2 z (\partial g g^{-1})_c (\bar{\partial} g g^{-1})^c.
\ee
The field $g$ takes values in a supergroup.

From the action we can calculate the classical currents associated to the invariance
of the theory under left multiplication of the field $g$ by a group element in $G_L$
and right multiplication by a group element in $G_R$.
 The classical equations of motion for the model read:
\begin{eqnarray}
\left( kf^2+1 \right) \bar{\partial} J^a + \left( kf^2-1 \right) \partial 
(g \bar{J} g^{-1})^a &=& 0,
\end{eqnarray}
where we have used the standard expressions for the left- and right-current
at the Wess-Zumino-Witten
 point\footnote{At the Wess-Zumino-Witten point, the parameters satisfy the equation: 
$1/f^2=k$.}:
\begin{align} 
J(z,\bar{z}) &= -k  \partial g g^{-1} \qquad\text{and}\qquad
\bar J(z,\bar z) = k g^{-1} \bar \partial g  \,.
\end{align}
The classical $G_L$ currents are given by:
\begin{align}\label{normeqn}
j_z &= -\frac{1}{2}\left(\frac{1}{f^2}+k\right) \partial g g^{-1}  = \frac{(1+kf^2)}{2kf^2} J \cr
j_{\bar{z}} &= -\frac{1}{2}\left(\frac{1}{f^2}-k\right) \bar{\partial} g g^{-1} = -\frac{(1-kf^2)}{2kf^2} (g \bar{J} g^{-1}).
\end{align}
At the Wess-Zumino-Witten point $f^2=1/k$ the $\bar{z}$-component of the
left-moving current becomes zero. As a consequence, the $z$-component $J$
becomes holomorphic. A similar phenomenon happens at the other Wess-Zumino-Witten point 
($f^2=-1/k$) for the anti-holomorphic component $g\bar{J}g^{-1}$.
From now on, we will concentrate on the current $j$
associated to the left action of the group.
For future reference we note that the coefficients that relate the left current components to the 
derivative of the group element are (see section \ref{generic}):
\begin{align}
c_+ &= -\frac{(1+kf^2)}{2f^2}\qquad\text{and}\qquad c_- = -\frac{(1-kf^2)}{2f^2} \,.
\end{align}

\subsection{Exact perturbation theory}

Elegant arguments were given \cite{Bershadsky:1999hk} for the
exactness of low-order perturbation theory for the calculation of
various observables in the supergroup model on $PSL(n|n)$. In
particular, we will use these arguments to compute the (left)
current-current two-point function exactly to all orders in
perturbation theory using the free theory. Similarly we also compute
the current three-point functions to all orders using perturbation
theory up to first order in the structure constants. Below, we
summarize some important facts that lead to these results \cite{Bershadsky:1999hk}.

The argument is essentially based on the special feature of the
Lagrangian that all interaction vertices are proportional to (powers
of) the structure constants as well as certain properties of the Lie
superalgebra, which we now list:

\begin{itemize}
\item If structure constants are doubly contracted, the result is proportional to the
Killing form, which is zero for the supergroups under consideration.

\item The only invariant three-tensor is proportional to the structure
  constant and the only invariant two-tensor is the invariant
  metric\footnote{These and the following are statements taken from
    \cite{Bershadsky:1999hk}. A detailed proof is 
lacking.  The
    condition of the uniqueness of the three-tensor can be relaxed to
    the condition that any invariant three-tensor contracted with the
    structure constants vanishes, which is a statement that has been
    proven in detail in an appendix to \cite{Quella:2007sg} for the
    particular case of 
the $psl(2|2)$ Lie superalgebra. }.


\item Traceless invariant $4$-tensors made of structure constants
and the invariant metric give zero when contracted with the
  structure constants over two indices \cite{Bershadsky:1999hk}.

\end{itemize} 

Using all these facts, and by using a pictorial representation of the
correlation functions, one can show that vacuum diagrams with at least
one interaction vertex all vanish and that group invariant correlation
functions can be computed in the free theory. However, in order to
compute the OPEs in the theory of interest, we have to calculate the
$2$ and $3$-point functions of the right-invariant currents $J^a$ and
$(g \bar J g^{-1})^a$, which are not fully invariant under the group
action.

In \cite{Bershadsky:1999hk}, it is shown that a correlation function
that is invariant under only the right group action, and which is a
two-tensor under the left group action (or vice versa), can be
computed by setting all structure constants to zero. Similarly, a
correlation function that is invariant under the right group action,
and a three-tensor under the left group action can be computed by
taking into account only contributions with at most one structure
constant. We will present the argument for the simplicity of the
$3$-point function in the next section.

In the following we also find it instructive to compute a four-point
function to second order in the structure constants. In order to
perform these calculations it is useful to expand the various terms in
the action as well as the currents to second order in the structure
constants.

\subsection*{Ingredients of perturbation theory}
We gather all our ingredients expanded to second order in the structure
constants. We use the conventions $g=e^A$ and $A=iA_a t^a$. For the left current
components we obtain:
\begin{eqnarray}\label{expCurrentA}
  J &=& - k \partial g g^{-1} =-k (\partial A+\frac{1}{2} [A,\partial A ]+
\frac{1}{3!} [ A, [ A , \partial A ] ]+ O(f^3))
  \nonumber \\
  &=& -k i (\partial A_a - \frac{{f^{bc}}_a}{2} A_b \partial A_c + 
\frac{1}{6} {f^a}_{bc} {f^c}_{de} A^b A^d \partial A^e+O(f^3)) t^a
  \nonumber \\
  g \bar{J} g^{-1} &=& k \bar{\partial} g g^{-1}
  = k  (\bar{\partial} A+\frac{1}{2} [A,\bar{\partial} A]+
\frac{1}{3!} [ A, [ A , \bar{\partial} A ] ]+ O(f^3))
  \nonumber \\
  &=& +ki (\bar{\partial} A_a - \frac{{f^{bc}}_a}{2} A_b \bar{\partial} A_c + 
\frac{1}{6} {f^a}_{bc} {f^c}_{de} A^b A^d \bar{\partial} A^e +O(f^3)) t^a,
\end{eqnarray}
where $O(f^3)$ indicates terms of third order or higher in the structure constants.
The kinetic term and the Wess-Zumino term become:
\begin{eqnarray}\label{expActionA}
S_{kin} &=& \frac{1}{4 \pi f^2}
\int d^2 z ( \partial A^a \bar{\partial} A_a -
\frac{1}{12} {f^a}_{bc} f_{aij} A^b \partial A^c A^i \bar{\partial} A^j
+ O(f^4))
\nonumber \\
S_{WZ} &=&  -\frac{k }{12 \pi} 
\int_{C} d^2 z    
 f^{abc} A_a  \partial A_b
  \bar{\partial} A_c + O(f^3).
\end{eqnarray}
The quadratic terms in the action give rise to the free propagators:
\begin{eqnarray}\label{freeProp}
A_a(z,\bar{z}) A_b(w,\bar{w}) &=& - f^2 \kappa_{ab} \log \mu^2 |z-w|^2,
\end{eqnarray}
where $\mu$ is an infrared regulator.

\subsection*{Two-point functions}

Consider the Feynman diagrams with two external lines and pull out a
structure constant where the external line enters. The rest of the
diagram is now a blob with three external lines, with two of them
contracted with the structure constant (the interaction
strength). Now, from a group theoretic perspective, the three-spoked
blob is also an invariant $3$-tensor, which, by the properties
itemized at the beginning of the section, is proportional to the
structure constant. As a result, the whole graph is proportional to
the metric times the Killing form: $f_{abc}f_d^{bc} =
2\check{h} g_{ad}$, which vanishes for the supergroups under
consideration.

The two-point functions are therefore perturbatively exact when computed by setting all structure constants to zero and follows directly from the free propagator \eqref{freeProp}:
\begin{eqnarray}
\langle J^a (z,\bar{z}) J^b (w,\bar{w}) \rangle
&=& \frac{f^2 k^2 \kappa^{ab}}{(z-w)^2}
\nonumber \\
\langle (g \bar{J} g^{-1})^a (z,\bar{z})  (g \bar{J} g^{-1})^b (w,\bar{w}) \rangle
&=& \frac{f^2 k^2\kappa^{ab}}{(\bar{z}-\bar{w})^2}
\nonumber \\
\langle J^a (z,\bar{z}) (g \bar{J} g^{-1})^b (w,\bar{w}) \rangle
&=& 2 \pi f^2 k^2\kappa^{ab} \delta^{(2)}(z-w). 
\end{eqnarray}

\subsection*{Three-point functions}

Consider the Feynman diagrams that contribute to 
$$
\langle J^a(z,\zbar) J^b(w,\wbar) J^c(x,\xbar)\rangle \,.
$$
In their evaluation, there are strucutre constants coming both from
the expansion of the currents in \eqref{expCurrentA} and also from the
interaction vertices.  We would like to argue that only those diagrams
which contain a single structure constant contribute. In order to show
this, consider pulling out one structure constant out of the 
vertex where the external line enters. The rest of the diagram can be
thought of as a blob with four external lines and which has the group
structure of a rank 4 invariant tensor. Contracting two of its
indices, the resulting graph contains a 
structure constant
inside and vanishes, following the same argument that allows to compute the two-point functions by setting all structure constants to zero.

We have now shown that the group structure of the four-spoked blob is that of a traceless rank 4 tensor. The full Feynman graph is evaluated by contracting a structure constant with this traceless rank-4 tensor. Using the special properties of the Lie superalgebra of $PSL(n|n)$ we listed earlier in the section, it is clear that such a term evaluates to zero. Thus, the three-point functions are perturbatively exact at first order in the structure constants. There are two non-trivial contributions to this calculation. We have one contribution coming from the term proportional to the structure constants in the expansion \eqref{expCurrentA} of the current components, and one from the first order Wess-Zumino interaction \eqref{expActionA}.

Let us compute the first contribution for the $JJJ$ three-point function in some detail:
\begin{align}
\langle  J^a (z,\bar{z}) J^b (w,\bar{w}) J^c(x,\bar{x}) \rangle_{1}
&= + i k^3 \langle (\partial A^a (z,\bar{z})- \frac{1}{2} f^{dea}
A_d \partial A_e)\cr
&    (\partial A^b - \frac{1}{2} f^{deb}
A_d \partial A_e)  (\partial A^c - \frac{1}{2} f^{dec}
A_d \partial A_e) \rangle
\cr
&= -i k^3 f^4 \frac{1}{2} ((+)\frac{ f^{abc} }{(z-x)(w-x)^2}
+ (+) \frac{ f^{bac} }{(w-x)(z-x)^2} + cyclic ) \cr
&= - i \frac{1}{2} k^3 f^4 f^{abc} \frac{z-w}{(z-x)^2(w-x)^2} + cyclic
\cr
&= - i  \frac{3}{2} k^3 f^4 f^{abc} \frac{1}{(z-w)(w-x) (x-z) } \,.
\end{align}
The Wess-Zumino contribution is\footnote{A minus sign arises from expanding $e^{-S}$ to first order.}:
\begin{align}
\left\langle  J^a (z,\bar{z}) J^b (w,\bar{w}) J^c(x,\bar{x}) \right\rangle_{2}
&= +i k^4 \left\langle \partial A^a (z,\bar{z}) \partial A^b(w,\bar{w}) \partial A^c(x,\bar{x})\times \right.
\cr
&\hspace{1in}\left. \frac{1}{12 \pi} \int_C d^2 y f^{deg} A_g \partial A_d \bar{\partial} A_e
(y, \bar{y}) \right\rangle
\cr
&=+ i k^4 f^6 \frac{1}{2} f^{abc} \frac{1}{(z-w)(w-x)(x-z)}.
\end{align}
Adding the two contributions we get the three-point function:
\begin{eqnarray}
\langle  J^a (z,\bar{z}) J^b (w,\bar{w}) J^c(x,\bar{x}) \rangle
&=&
- i \frac{1}{2} k^3 f^4 (3 -  k f^2) f^{abc}  \frac{1}{(z-w)(w-x)(x-z)}.
\end{eqnarray}
A quick check on the calculation is that it matches the known three-point function
at the Wess-Zumino-Witten point, where it can be evaluated using the holomorphy of the currents.
All other left current three-point functions can be computed analogously. They
are (up to contact terms):
\begin{align}\label{exactJJJ}
\langle J^a (z,\bar{z}) J^b (w,\bar{w}) J^c (x,\bar{x}) \rangle &=
 \left( \frac{3}{2}- \frac{kf^2}{2}\right) \frac{- i k^3 f^4 f^{abc}}{(z-w)(w-x)(x-z)} \cr
\langle J^a (z,\bar{z}) J^b (w,\bar{w})(g \bar{J} g^{-1})^c (x,\bar{x}) \rangle &=
  \left( \frac{1}{2}- \frac{kf^2}{2}\right) 
\frac{- i k^3 f^4 f^{abc}(\bar{z}-\bar{w})}{(z-w)^2(\bar{x}-\bar{w})(\bar{x}-\bar{z})}  
\cr
\langle (g \bar{J} g^{-1})^a (z,\bar{z}) (g \bar{J} g^{-1})^b (w,\bar{w}) J^c (x,\bar{x}) \rangle &=
 \left( \frac{1}{2}+ \frac{kf^2}{2}\right) 
\frac{+ i k^3 f^4 f^{abc} (z-w)}{(\bar{z}-\bar{w})^2(x-w)(x-z)} 
\cr
\langle (g \bar{J} g^{-1})^a (z,\bar{z})(g \bar{J} g^{-1})^b (w,\bar{w})(g \bar{J} g^{-1})^c (x,\bar{x}) \rangle &= 
  \left( \frac{3}{2}+ \frac{kf^2}{2}\right) 
\frac{+ i k^3 f^4 f^{abc}}{(\bar{z}-\bar{w})(\bar{w}-\bar{x})(\bar{x}-\bar{z})} \,.
\end{align}
\subsection*{Coincidence limit and operator product expansions}
When we take coincidence limits of the three-point functions, we
expect to be able to replace the product of two operators by their
operator product expansion. Using the general form of the
current-current operator product expansions, and the exact two-point
functions, we can infer from the above three-point functions a
proposal for the current-current operator product expansions.  Up to
contact terms the two- and three-point functions can be reproduced in
their coincidence limits by the OPEs \footnote{From now on we will no
  longer always make explicit the fact that all the operators depend
  both on the holomorphic as well as the anti-holomorphic coordinate
  at a generic point in the moduli space.}:
\begin{align}\label{finiteJXOPE}
J^a (z,\bar{z}) J^b (w,\bar{w}) & \sim  \frac{k^2 f^2 \kappa^{ab}}{(z-w)^2}
+ k f^2 ( \frac{3}{2}-\frac{kf^2}{2}) i {f^{ab}}_c \frac{J^c (w)}{z-w}\cr
&  + (-kf^2) ( \frac{1}{2} - \frac{kf^2}{2}) i {f^{ab}}_c \frac{\bar{z}-\bar{w}}{(z-w)^2}
(g \bar{J} g^{-1})^c(w) +\ldots \cr
J^a (z,\bar{z}) (g \bar{J} g^{-1})^b (w,\bar{w}) & \sim 2 \pi {k^2 f^2}\kappa^{ab} \delta^{(2)}(z-w)
+ k f^2 ( \frac{1}{2}+\frac{kf^2}{2}) i {f^{ab}}_c \frac{(g \bar{J} g^{-1})^c (w)}{z-w}\cr
& -kf^2 ( \frac{1}{2} - \frac{kf^2}{2}) i {f^{ab}}_c \frac{1}{\bar{z}-\bar{w}}
J^c (w) + \ldots \cr
(g \bar{J} g^{-1})^a (z,\bar{z}) (g \bar{J} g^{-1})^b (w,\bar{w}) & \sim  \frac{k^2 f^2\kappa^{ab}}{(\bar{z}-\bar{w})^2}
- k f^2 ( \frac{3}{2}+\frac{kf^2}{2}) i {f^{ab}}_c \frac{(g \bar{J} g^{-1})^c (w)}{\bar{z}-\bar{w}} \cr
& + kf^2 ( \frac{1}{2} + \frac{kf^2}{2}) i {f^{ab}}_c \frac{z-w}{(\bar{z}-\bar{w})^2}
J^c(w) + \ldots
\end{align}
When we normalize the currents as in \eqref{normeqn} to agree with section \ref{generic}, we find the following OPEs:
\begin{align}
j^a_z (z) j_{w}^b (w) & \sim  \frac{(1+kf^2)^2\kappa^{ab}}{4 f^2(z-w)^2}
+  \frac{i}{4} (1+kf^2)(3-kf^2) {f^{ab}}_c \frac{j^c_w (w)}{z-w}\cr  
&\hspace{2cm}+\frac{i}{4}(1+kf^2)^2{f^{ab}}_c \frac{\bar{z}-\bar{w}}{(z-w)^2} j^c_{\bar{w}}(w) + \ldots \cr
j^a_{\bar{z}} (z) j^b_{\bar{w}} (w) & \sim  \frac{(1-kf^2)^2\kappa^{ab}}{4f^2\, (\bar{z}-\bar{w})^2}
+ \frac{i}{4} (1-kf^2) (3+kf^2) \frac{ {f^{ab}}_c\, j^c_{\bar{w}} (w)}{\bar{z}-\bar{w}} \cr
&\hspace{2cm}+ \frac{i}{4} (1-kf^2)^2  \frac{(z-w)\,  {f^{ab}}_c\, j^c_w(w)}{(\bar{z}-\bar{w})^2}  + \ldots\cr
j^a_z (z) j^b_{\bar{w}} (w) & \sim  - \frac{2\pi}{4f^2} (1+kf^2)(1-kf^2) \kappa^{ab}\delta^{(2)}(z-w)
+ \frac{i(1+kf^2)^2}{4}  \frac{{f^{ab}}_c\, j^c_{\bar{w}} (w)}{z-w} \cr
&\hspace{2cm} +\frac{i(1 -kf^2)^2}{4}  \frac{{f^{ab}}_c\, j^c_w (w) }{\bar{z}-\bar{w}}+\ldots 
\end{align}
We can read from these formulas the coefficients of the generic current algebra \eqref{euclidOPEs}:
\begin{align}\label{cisupergroup}
c_1 &= \frac{(1+kf^2)^2}{4 f^2} \qquad c_2 =  \frac{i}{4}(1+kf^2)( 3-kf^2)  \cr
c_3 &=  \frac{(1-kf^2)^2}{4f^2}\qquad c_4 =  \frac{i}{4}(1-kf^2)(3+kf^2) \cr
&\text{and}\quad g = \frac{i}{2}(1+kf^2)(1-kf^2) \,. 
\end{align}
We note that the coefficients automatically satisfy the extra constraints \eqref{MCone} and \eqref{MCtwo} one gets by requiring consistency of the current algebra with the Maurer-Cartan equation. 

\subsection*{A four-point function}
From the three-point functions, we conclude that the coefficient $g$
is non-zero. We have argued in section \ref{generic} that that is
associated to the appearance of logarithms in the regular term of the
$j_z j_{\bar{z}}$ operator product expansion. We would like to check the coefficient of
the logarithm more directly in a perturbative calculation. For that purpose it
 is sufficient
to study a four-point function at second order in the structure constants. The computation will be exact at that order. In particular
we want to compute the four-point function:
\begin{eqnarray}
\langle J^{[a} (z) (g \bar{J} g^{-1})^{b]}(w) J^c (x) (g \bar{J} g^{-1})^d(y) \rangle_{O(f^2)},
\end{eqnarray}
at second order in the structure constants, and in the $z
\rightarrow w$ limit.  We anti-symmetrized in the $a$ and $b$ indices
(and weighted each term with a factor $1/2$). In the coincidence
limit, logarithms appear in the regular terms in the OPE between $J$
and $(g \bar{J} g^{-1})$ and they will give non-zero contribution to
the four-point functions. In our calculation we focus on 
the terms proportional to $\log |z-w|^2$ (and which are not contact terms).
 
We distinghuish the following contributions at this order.
We can expand a single current to second order, and compute in the free theory.
We can expand two currents to first order and compute in the free theory.
We can add one linear Wess-Zumino interaction term and expand one current to first order.
Or we can add two linear Wess-Zumino interaction terms and take only the leading
terms in the currents.
Finally, we can add one quadratic principal chiral model interaction term, and treat the currents
at zeroth order.

We found the following results. The second order term in a current cannot give
rise to logarithmic contributions. Two currents at first order can be contracted
to give a logarithm. It is easy to see that there are few contributions to the terms
of interest, and they give: 
\begin{eqnarray}
- \frac{k^4 f^6}{8} f^{abe} {f^{cd}}_e \log \mu^2 |z-w|^2 \frac{1}{(z-x)^2 
(\bar{w}-\bar{y})^2}
\end{eqnarray}
The term linear in the Wess-Zumino interaction term gives no contribution.
The term arising from the quartic interaction term in the principal chiral model 
doubles the previous non-zero term.
The Wess-Zumino term squared gives a contribution of a different type equal to:
\begin{eqnarray}
+ \frac{k^6 f^{10}}{4} \log \mu^2 |z-w|^2f^{abe} {f^{cd}}_e \frac{1}{(z-x)^2 
(\bar{w}-\bar{y})^2}.
\end{eqnarray}
The latter contribution is the hardest to calculate. It consists of
the order of $216$ free field contractions, which exhibit a lot of
symmetries. Some logarithmic terms in the double integrals over the points
of interaction need to be
evaluated, but all of these integrals are straightforwardly performed
using partial integrations and other elementary techniques. The tedious but elementary
calculation leads to the above result. In total
we get, at second order in the structure constants, and only regarding
the logarithmic contribution in $z,w$:
\begin{multline}
\langle J^{[a} (z) (g \bar{J} g^{-1})^{b]}(w) J^c (x) (g \bar{J} g^{-1})^d(y) \rangle_{O(f^2),log}
=\cr
-\frac{1}{4} k^4 f^6 (1- k^2 f^{4}) f^{abe} {f^{cd}}_e \log \mu^2 |z-w|^2 \frac{1}{(z-x)^2 
(\bar{w}-\bar{y})^2} \,.
\end{multline}
Using the relative normalization between the $J$'s and the $j$'s, we find that
\begin{multline}\label{logContrib}
\langle j_z^{[a} (z) j_{\bar w}^{b]}(w) j_{x}^c (x) j_{\bar y}^d(y) \rangle_{O(f^2),log}= 
\cr
-\frac{1}{64 f^2}(1+kf^2)^2(1-kf^2)(1-k^2f^4)f^{abe} {f^{cd}}_e \log \mu^2 |z-w|^2 \frac{1}{(z-x)^2 
(\bar{w}-\bar{y})^2} \,.
\end{multline}
Let us see how to use this result to check the coefficient $g$ in the operator product expansion.
From the  expressions for the operator product algebra \eqref{euclidOPEs} and from the exact
three-point functions \eqref{exactJJJ}, we find that the logarithmic term in the normalized four-point function in the coincidence limit 
$z \rightarrow w$ is:
\begin{eqnarray}
\langle j_z^a j_{\bar{w}}^b j_x^c j_{\bar{y}}^d \rangle
& \approx &
- \frac{g}{4} {f^{ab}}_e \log \mu^2 |z-w|^2 
\langle (\partial j^e_{\bar{w}}
- \bar{\partial} j_w^e) j_x^c j_{\bar{y}}^d \rangle
\nonumber \\
& \approx& + \frac{g}{4} \frac{1}{8k^3 f^6} (1+kf^2)(1-kf^2) {f^{ab}}_e \log \mu^2 |z-w|^2 
\nonumber \\
& & 
(-\partial_w (1-kf^2)  i k^3 f^4 f^{edc} \frac{1}{2} (1+kf^2)
 \frac{w-y}{(\bar{w}-\bar{y})^2 (x-w)(x-y)}
\nonumber \\
& & 
- \bar{\partial}_{w} (1+kf^2) f^{ecd}(-) i k^3 f^4 \frac{1}{2} (1-kf^2)
\frac{\bar{w}-\bar{x}}{(w-x)^2 (\bar{y}-\bar{w})(\bar{y}-\bar{x})})
\nonumber \\
& \approx& +i \frac{g}{32f^2}(1+kf^2)^2(1-kf^2)^2 {f^{ab}}_e f^{ecd} 
\nonumber \\
& &
\log \mu^2 |z-w|^2  \frac{1}{(w-x)^2 (\bar{w}-\bar{y})^2}.
\end{eqnarray}
We recall the value for $g$:
\be
g = \frac{i}{2} (1+kf^2) (1-kf^2),
\ee
so the operator algebra and the three-point functions predict:
\begin{eqnarray}
\langle j_z^a j_{\bar{w}}^b j_x^c j_{\bar{y}}^d \rangle
& \approx & - \frac{1}{64 f^2} (1+kf^2)^2(1-kf^2)^2
(1-k^2 f^4) {f^{ab}}_e f^{ecd} 
\nonumber \\
& &
\log \mu^2 |z-w|^2  \frac{1}{(w-x)^2 (\bar{w}-\bar{y})^2}.
\end{eqnarray}
The prediction is matched by our perturbative calculation of the
four-point function.  Moreover, since the coefficient $g$ is fixed to
all orders by the calculation of the three-point function, our result
at second order in the structure constants is exact. The calculation
is a good consistency check on the correlators and operator product
expansions. The full four-point function is a function of the cross
ratio of the four insertion points, in which the regulator $\mu$ drops
out. We note also that the appearance of logarithms in four-point
functions of operators that differ by an integer in their conformal
dimension is generic. In our case, the scale $\mu$ must appear in the
operator product expansion because we lifted space-time fermionic
zeromodes \cite{Bershadsky:1999hk}. These in turn are linked to the 
non-diagonalizable nature of the scaling operator in sigma-models
on supergroups \cite{Rozansky:1992rx}.

\subsection{Summary of the current algebra}

We summarize the current algebra for the left group action:
\begin{align}\label{defalgebralogs}
j^a_z (z) j_{z}^b (w) & =  \frac{(1+kf^2)^2 \kappa^{ab}}{4 f^2 (z-w)^2}
+  \frac{i}{4}(1+kf^2)(3-kf^2) \frac{{f^{ab}}_c \, j^c_z (w)}{z-w} \cr
& +\frac{i}{4}(1+kf^2)^2 \frac{\bar{z}-\bar{w}}{(z-w)^2}
 {f^{ab}}_c\, j^c_{\bar{z}}(w) + :j^a_z j_{z}^b: (w)\cr
j^a_z (z) j^b_{\bar{z}} (w) & =  - \frac{\pi}{4 f^2}(1+kf^2)(1-kf^2) \kappa^{ab}\delta^{(2)}(z-w)
+  \frac{i (1+kf^2)^2}{4}  \frac{{f^{ab}}_c\, j^c_{\bar{z}} (w)}{z-w}\cr
& +\frac{i (1 -kf^2)^2}{4}  \frac{{f^{ab}}_c\, j^c_z (w)}{\bar{z}-\bar{w}}
 - \frac{i}{8}(1-k^2f^4) {f^{ab}}_c \log|z-w|^2 \left( \partial j^c_{\bar z}(w) - \bar \partial j^c_z(w) \right)\cr
& + :j^a_z j^b_{\bar{z}}: (w) \cr
j^a_{\bar{z}} (z) j^b_{\bar{z}} (w) & = 
 \frac{(1-kf^2)^2\kappa^{ab}}{4f^2 (\bar{z}-\bar{w})^2}
+  \frac{i}{4}(1-kf^2) ( 3+kf^2)  \frac{ {f^{ab}}_c\, j^c_{\bar{w}} (w)}{\bar{z}-\bar{w}} \cr
&+\frac{i (1-kf^2)^2}{4} \frac{z-w}{(\bar{z}-\bar{w})^2}
 {f^{ab}}_c\, j^c(w) + :j^a_{\bar{z}} j^b_{\bar{z}}: (w) \,.
\end{align}
The current algebra for the right group action can be obtained through
the combined operation $g \rightarrow g^{-1}$ and worldsheet parity $P$, which
is a symmetry of the model

\section{Conformal perturbation theory}

\label{firstorder}
In this section we study Wess-Zumino-Witten models with a perturbed
kinetic term, both for its intrinsic interest and as a tractable
example of conformal perturbation theory. For a general group
manifold, the deformed model becomes non-conformal. For supergroup
manifolds with vanishing Killing form, the models remain conformal.
In the earlier sections, we have computed, using the exact two- and
three-point functions, the current algebra of the deformed theory. In
this section, we re-derive  these results using conventional conformal perturbation
theory. This will be a consistency check of the deformed conformal
current algebra we have obtained in equation \eqref{defalgebralogs}.

\subsection{The current algebra in the Wess-Zumino-Witten model}
We first review the chiral
current algebra of the Wess-Zumino-Witten model. We recall the action
\be S_{WZW} = \frac{k}{ 16 \pi}\int d^2 x Tr'[- \partial^\mu g^{-1}
\partial_\mu g] + k \Gamma \ee
where $\Gamma$ is the Wess-Zumino term, and the field $g(z,\bar z)$ takes values in a (super)group $G$. The model has a global  $G_L \times G_R$ invariance by left and right multiplication of the group element. The currents associated to these symmetries are (in complex coordinates):
\begin{align} 
J(z) &= -k  \partial g g^{-1} \qquad\text{and}\qquad
\bar J(\bar z) = k g^{-1} \bar \partial g  \,.
\end{align}
The right-invariant current $J(z)$ is holomorphic and generates the left-action of the group $G_L$.  The left-invariant anti-holomorphic current $\bar{J}$ generates the right translation by a group element. The components of the current $J$ satisfy the OPE: 
\be J^a(z) J^b(w) \sim \frac{k \kappa^{ab}}{(z-w)^2} + i f^{ab} {}_c \frac{J^c(w)}{z-w},
\ee 
and the components of the current $\bar J (\bar z)$ satisfy the same
OPE, with anti-holomorphic coordinates instead of holomorphic ones. In
particular, in our conventions the pole term keeps the same sign.
These currents generate a large chiral affine current algebra whose existence is
useful in solving the model via the Knizhnik-Zamolodchikov equations.

\subsection{Perturbation of the kinetic term: classical analysis}

We are interested in the following marginal
deformation of the Wess-Zumino-Witten model: 
\be S = S_{WZW} +\frac{\lambda}{4\pi k} \int d^2z\, \Phi(z,\bar z)\,. 
\ee 
where
\be\label{defPhi}
\Phi = \frac{1}{2} (:J^c (g \bar{J} g^{-1})_c: + :(g \bar{J} g^{-1})_c J^c:)\,.
\ee
In other words, we perturb the kinetic term by multiplying it with a
factor $1+\lambda$.  Comparing the action with the action in the
earlier section, we find that $\lambda$ is related to the kinetic
coefficient $f$ defined in the previous section by the relation
\be\label{fklambda}
\frac{1}{f^2} = k(1+\lambda) \,.
\ee
We note that, analogous to the composite operator that appeared in the
Maurer-Cartan equation, we have chosen a symmetric combination of the
product of $J$ and $g \bar{J} g^{-1}$ operators to represent the
marginal operator in the quantum theory.

\subsection{The current-current operator product expansions}

In this subsection we compute the correction to the holomorphic
current-current operator product expansion induced by the perturbation
of the kinetic term of the Wess-Zumino-Witten model for a simple (super)
Lie algebra.
In order to perform the calculation we require the OPEs between the
currents $J$ and $g \bar J g^{-1}$ at the WZW point. These are
obtained by requiring that the Maurer-Cartan equation holds in the
quantum WZW model, as shown in Appendix \ref{WZWOPE}: we compute the
OPE of the current $J$ with the Maurer-Cartan equation for a generic
value of the dual Coxeter number and demand that it vanish. This
constraint leads to the operator product expansion
\be 
J^a(z) (g \bar J g^{-1})^b(w,\bar w) = 
2\pi k \kappa^{ab} \delta^{(2)}(z-w) + i {f^{ab}}_c \frac{(g \bar J g^{-1})^c(w,\bar w)}{z-w} 
+ :J^a (g \bar J g^{-1})^b:(w,\bar w) \,. 
\ee
A similar demand on contact terms and the most singular terms in the OPE of $(g \bar{J} g^{-1})$
with the Maurer-Cartan equation leads to the OPE
\begin{multline} (g \bar J g^{-1})^a(z,\bar z)(g \bar J g^{-1})^b(w,\bar w) = \frac{k \kappa^{ab}}{(\bar z - \bar w)^2} + i {f^{ab}}_c \frac{J^c(w)(z-w)}{(\bar z - \bar w)^2} - 2i {f^{ab}}_c \frac{(g \bar J g^{-1})^c(w,\bar w)}{\bar z - \bar w} \cr 
+ :(g \bar J g^{-1})^a(g \bar J g^{-1})^b:(w,\bar w)\,. 
\end{multline}
 A general
discussion of higher order corrections to operator product expansions
is given in Appendix \ref{perturbOPE}. Here we focus on applying
the discussion to the case of a supergroup with vanishing Killing form.
We compute the corrections induced by
the exactly marginal perturbation to the $J^a J^b$ OPE.
The $n$th
order correction is denoted by
\be 
(JJ)^{ab}_n(z-w,w) = \left[ J^a(z,\bar z) \frac{(-\lambda)^n}{(4\pi k)^n n!} \prod_{i=1}^n \int d^2x_i \Phi(x_i, \bar x_i) \right] J^b(w,\bar w) 
\ee
where the square bracket means that we have to contract $J^a(z,\bar z)$ with all the integrated operators before we contract it with $J^b(w,\bar w)$. We define $H^a_n(z,\bar z)$ to be this complete contraction:
\be \label{H}
H^a_n(z,\bar z) =  \left[ J^a(z,\bar z)\frac{1}{(4\pi k)^n n!} \prod_{i=1}^n \int d^2 x_i \Phi(x_i, \bar x_i) \right] \,.
\ee
One can similarly define another contraction, with $J$ replaced by the current $(g \bar{J} g^{-1})^a$:
\be\label{Hbar}
\bar{H}^a_n(z,\bar z) =  \left[ (g \bar{J} g^{-1})^a(z,\bar z)\frac{1}{(4\pi k)^n n!} \prod_{i=1}^n \int d^2 x_i \Phi(x_i, \bar x_i) \right] \,.
\ee
The basic building blocks we need to carry out this computation are
the OPE of the currents $J^a$ and $(g \bar{J} g^{-1})^a$ with the
marginal operator $\Phi$. As we will see, once these OPEs are
obtained, the $n$th order correction can be obtained by a process of
iteration. Here, we list the two OPEs of interest and refer the
reader to Appendix \ref{WZWOPE} for details. 
\begin{align}\label{JPhiXPhi}
J^a (w) \int d^2 x \Phi (x, \bar{x}) &\sim  k \int d^2 x \frac{ (g \bar{J} g^{-1})^a(w,\wbar)}{(w-x)^2} + 2 \pi k J^a (w)\cr
(g \bar{J} g^{-1})^a(w,\bar w)  \int d^2 x  \Phi(x, \bar x)
&\sim 6\pi k (g \bar{J} g^{-1})^a(w,\bar w) -  k \int d^2 x \frac{J^a(x,\bar x)}{(\bar w- \bar x)^2} \,. 
\end{align}
With these basic OPEs, let us contract the current with one of the integrated marginal operators:
\begin{align}  
H^a_n(z,\bar z) &= \frac{n}{(4\pi k)^n n!} \int d^2 x \left (\frac{k (g \bar{J} g^{-1})^a(x,\bar x)}{(z-x)^2} + 2\pi k \delta^{(2)}(z-x) J^a(x,\bar x) \right)\prod_{i=1}^{n-1} \int d^2 x_i \Phi(x_i, \bar x^i) \cr
&=  \frac{1}{4\pi} \int d^2 x\left( \frac{\bar H^a_{n-1}(x,\bar x)}{(z-x)^2} \right)
+  \frac{1}{2} H^a_{n-1}(z,\bar z)
\end{align}
One can do a similar operation on $\bar H_n^a$ and we get
\begin{align}  
\bar H^a_n(z,\bar z) &=  \frac{n}{(4\pi k)^n n!}\left[
6\pi k (g \bar{J} g^{-1})^a(z,\bar z) - \int d^2 x\frac{k J^a(x,\bar x)}{(\bar z - \bar x)^2} \right] 
\prod_{i=1}^{n-1} \int d^2 x_i \Phi(x_i, \bar x^i) \cr
&= \frac{3}{2} \bar H^a_{n-1}(z,\bar z)  - \frac{1}{4\pi} \int d^2 x\left( \frac{ H^a_{n-1}(x,\bar x)}{(\bar z-\bar x)^2} \right) \,.
\end{align}
These are coupled recursion relations for $H_n^a$ and $\bar H_n^a$ subject to the initial conditions:
\beq H^a_0(z,\bar z) &=& J^a(z,\bar z) \nonumber\\
\bar H^a_0(z,\bar z) &=& (g \bar{J} g^{-1})^a(z,\bar z)
\eeq
These recursion relations have the following solutions:
 \begin{align} 
 H^a_n(z,\bar z) &= \left( 1-\frac{n}{2} \right)  J^a(z,\bar z) + \frac{n}{2}\frac{1}{2\pi} \int d^2x \frac{(g \bar{J} g^{-1})^a(x,\bar x)}{(z-x)^2} \cr
 \bar H^a_n(z,\bar z) &= \left( 1+\frac{n}{2} \right) (g \bar{J} g^{-1})^a(z,\bar z) 
-\frac{n}{2} \frac{1}{2\pi} \int d^2x \frac{J^a(x,\bar x)}{(\bar z-\bar x)^2}\,.
 \end{align}
In particular we deduce that
\begin{align}
(JJ)^{ab}_n(z-w,w) &=  (-\lambda)^n J^b(w,\bar w)
\left[ \left( 1-\frac{n}{2} \right) J^a(z,\bar z) + \frac{n}{2} \frac{1}{2\pi} \int d^2x \frac{(g \bar{J} g^{-1})^a(x,\bar x)}{(z-x)^2}
\right]\cr
&=   (-\lambda)^n \left[ \left( 1-\frac{n}{2} \right) \left(
\frac{k \kappa^{ab}}{(z-w)^2} + i f^{abc} \frac{J^c(w)}{z-w} \right) \right. \cr
& + \left . \frac{n}{2}\frac{1}{2\pi} \int d^2x \frac{1}{(z-x)^2} \left(
2\pi k \kappa^{ab} \delta^{(2)}(z-x) + i f^{abc}\frac{(g \bar{J} g^{-1})^c(w,\bar w)}{x-w}
\right)\right] \cr
&=   (-\lambda)^n \left[
 \frac{k \kappa^{ab}}{(z-w)^2} + i f^{abc} \left( 1-\frac{n}{2} \right) \frac{J^c(w)}{z-w} \right. \cr
 &\hspace{1.5in}\left.  +  i \frac{n}{2}f^{abc}(g \bar{J} g^{-1})^c(w,\bar w)  \frac{\bar z - \bar w}{(z-w)^2} + \ldots \right]\,. 
\end{align}
We can now sum the perturbative series in $\lambda$. We get the OPE in the perturbed theory: 
\begin{align} 
J^a&(z,\bar z) J^b(w,\bar w) =\sum_{n =0}^{\infty} (-\lambda)^n (JJ)^{ab}_n(z-w,w) \cr
&=\frac{1}{1+\lambda} \frac{k \kappa^{ab}}{(z-w)^2}
+ \frac{2+3\lambda}{2(1+\lambda)^2} i f^{abc} \frac{J^c(w)}{z-w}
- \frac{\lambda}{2(1+\lambda)^2}    i f^{abc}(g \bar{J} g^{-1})^c(w,\bar w)  \frac{\bar z - \bar w}{(z-w)^2} \nonumber
\end{align}
Using the map $(kf^2)^{-1} =1+\lambda$, one can check that this coincides with the OPE  in equation \eqref{finiteJXOPE}.
With the same techniques we can also compute the corrections to the $J^a (g \bar J g^{-1})^b$ and $(g \bar J g^{-1})^a(g \bar J g^{-1})^b$ OPEs. We get the results
\begin{multline} 
J^a(z,\bar z) (g \bar{J} g^{-1})^b (w, \bar w) = \frac{1}{1+\lambda} 2 \pi k \kappa^{ab} \delta^{(2)}(z-w) 
+ i f^{abc} \frac{2+\lambda}{2(1+\lambda)^2} \frac{(g \bar{J} g^{-1})^c(w,\bar w)}{z-w}\cr
- i f^{abc} \frac{\lambda}{2(1+\lambda)^2} \frac{J^c(w)}{\bar z-\bar w} +\ldots
\end{multline}
\begin{multline} 
  (g \bar{J} g^{-1})^a(z,\bar z) (g \bar{J} g^{-1})^b (w, \bar w) =
  \frac{1}{1+\lambda} \frac{k \kappa^{ab}}{(\zbar-\wbar)^2}
  -\frac{4+3\lambda}{2(1+\lambda)^2} \frac{i f^{abc}(g \bar{J}
    g^{-1})^c(w,\bar w}{\bar z-\bar w} \cr +
  \frac{2+\lambda}{2(1+\lambda)^2} \frac{(z-w)i f^{abc} J^c(w)}{(\bar
    z-\bar w)^2}+...
\end{multline} 
Again this matches with the OPE obtained in equation \eqref{finiteJXOPE}.

\subsubsection*{Summary}

In this section, we have shown by resumming conformal perturbation theory
that the deformed current algebra we obtain
this way  matches the algebra obtained in section
\ref{correlators} through the calculation of 2- and 3-point functions to all
orders in perturbation theory.

\section{The current algebra on the cylinder}
\label{cylinder}
In this section we consider the sigma-model on a cylinder and Fourier decompose
the current algebra. The representation in terms of Fourier modes is often
more conventient. To put the algebra on a cylinder, we first
compute the operator algebra on the plane, and then compactify the
plane. We consider the complex plane $z=\sigma-i \tau$ and consider
$\tau$ as time and $\sigma$ as the spatial coordinate. Denoting the currents as $j_{\mu}(\sigma,\tau)$, the commutator of equal-time operators is defined to be the limit of the difference of time-ordered operators (evaluated at $\tau=0$)
\be
{[} j_\mu^a (\sigma,0) , j_\nu^b(0,0) {]} = \lim_{\epsilon \rightarrow 0} (j_\mu^a (\sigma, i \epsilon) j_\nu^b(0,0) - j_\nu^b(0,i \epsilon) j_\mu^a(\sigma,0)) \,.
\ee
Using this definition, let us compute the commutators for the holomorphic component of the current (we suppress the $\tau=0$ argument within the currents in what follows):
\begin{align}
{[} j_z^a (\sigma) , j_z^b(0) {]}
&=  \lim_{\epsilon \rightarrow 0} \left\{
 \frac{c_1 \kappa^{ab}}{(\sigma-i \epsilon)^2} +  {f^{ab}}_c
(  \frac{c_2}{\sigma-i \epsilon} j_z^c(0) + (c_2-g)
\frac{\sigma+i \epsilon}{(\sigma-i \epsilon)^2} j_{\bar{z}}^c(0))\right. \cr
& \hspace{1.5cm}\left. - \frac{c_1 \kappa^{ab}}{(\sigma+i \epsilon)^2} -  {f^{ab}}_c
(\frac{c_2}{-\sigma-i \epsilon} j_z^c(\sigma) +  (c_2-g)
\frac{-\sigma+i \epsilon}{(\sigma+i \epsilon)^2} j_{\bar{z}}^c(\sigma))\right\}
+ \dots\cr
&= - 2 \pi i  c_1 \delta'(\sigma) \kappa^{ab}
+ 2 \pi i c_2 \delta(\sigma)  {f^{ab}}_c j^c_z(0)
+ 2 \pi i  (c_2-g)  \delta(\sigma) {f^{ab}}_c j^c_{\bar{z}} (0) + \dots
\end{align}
where we used
\begin{align}
\lim_{\epsilon \rightarrow 0} \frac{1}{\sigma- i \epsilon} -
\frac{1}{\sigma+ i \epsilon} &=  2 \pi i \delta(\sigma)\cr
\lim_{\epsilon \rightarrow 0} \frac{1}{(\sigma- i \epsilon)^2} -
\frac{1}{(\sigma+ i \epsilon)^2} &= - 2 \pi i \delta'(\sigma)\cr
\lim_{\epsilon \rightarrow 0} \frac{\sigma+i \epsilon}{(\sigma- i \epsilon)^2} -
\frac{\sigma- i \epsilon}{(\sigma+ i \epsilon)^2} &= 2 \pi i \delta(\sigma).
\end{align}
For other components we find:
\begin{align}
{[} j_{\bar{z}}^a (\sigma) , j_{\bar{z}}^b(0) {]}
&= + 2 \pi i c_3 \delta'(\sigma)\kappa^{ab} - 2 \pi i c_4\delta(\sigma) {f^{ab}}_c
j_{\bar{z}}^c(0)- 2 \pi i (c_4-g) \delta(\sigma)  {f^{ab}}_c j_{z}^c(0) \cr
{[} j_{z}^a (\sigma) , j_{\bar{z}}^b(0) {]}
&=- 2 \pi i  (c_4-g) \delta(\sigma) {f^{ab}}_c j_z^c(0) 
+  2 \pi i (c_2-g)   \delta(\sigma) {f^{ab}}_c j_{\bar{z}}^c(0) \,.
\end{align}
It is now straightforward to compactify $\sigma \equiv \sigma+2 \pi$ and
Fourier decompose the operator algebra on the cylinder using:
\begin{eqnarray}
j_z &=& +i \sum_{n \in Z} e^{-i n \sigma} j_{z,n}
\nonumber \\
j_{\bar{z}} &=& -i \sum_{n \in Z} e^{-i n \sigma} j_{\bar{z},n}
\nonumber \\
\delta(\sigma) &=& \frac{1}{2 \pi} \sum_{n \in Z} e^{i n \sigma}.
\end{eqnarray}
We find:
\begin{eqnarray}
{[} j_{z,n}^a , j_{z,m}^b {]} &=&
c_1 \kappa^{ab} n \delta_{n+m,0} +  c_2 {f^{ab}}_c j_{z,n+m}^c
-   (c_2-g) {f^{ab}}_c j_{\bar{z},n+m}^c
\nonumber \\
{[} j_{\bar{z},n}^a , j_{\bar{z},m}^b {]} &=&
-c_3 \kappa^{ab} n \delta_{n+m,0} + c_4 {f^{ab}}_c j_{\bar{z},n+m}^c
-   (c_4-g) {f^{ab}}_c j_{z,n+m}^c
\nonumber \\
{[} j_{z,n}^a , j_{\bar{z},m}^b {]} &=&
(c_4-g)  {f^{ab}}_c j_{z,n+m}^c +  (c_2-g)  {f^{ab}}_c  j_{\bar{z},n+m}^c.
\end{eqnarray}
We can check the validity of the Jacobi identity, which follows from
the validity of the Maurer-Cartan equation along with the Jacobi
identity for the Lie algebra of $G$.
\subsection*{Conserved charges}
We note that the current one-form satisfies the conservation equation
 $d \ast j =0$, and that therefore the integral of the time-component of the
current over the spatial circle is conserved in time. The corresponding charges
are easily determined to be the sum of the zero-modes of the current algebra.
They generate the Lie algebra of $G$. We recall that the group action
generated by these charges corresponds to the left group action $G_L$,
and that there is an analogous right group action $G_R$.
\subsection*{Kac-Moody subalgebra}

Let us consider the combination of the currents $j^a_{z}-j^a_{\bar z}$
and compute the commutation relations of its modes with themselves.
Using the above basic commutation relations, we find
\begin{align} 
[(j^a_{z,n}+j^a_{\bar z,n}), (j^b_{z,m}+j^b_{\bar z,m})] &= (c_1-c_3) \kappa^{ab} n
  \delta_{m+n,0}+(c_2+c_4-g){f^{ab}}_c (j^c_{z,n+m}+j^c_{\bar z,n+m})\,,
\end{align}
which is a Kac-Moody algebra at level 
\be
k^+=-\frac{c_1-c_3}{(c_2+c_4-g)^2}
\ee
as becomes manifest in terms of the rescaled currents
\be
{\cal J}^a = - i \frac{j^a_z - j^a_{\bar z}}{c_2+c_4-g} \,.
\ee
We observe that for the case of the supergroup considered in the
earlier section, substituting the values of the $c_i$ in
\eqref{cisupergroup}, we obtain a Kac-Moody algebra at level
$k^+=k$, with the currents taking the simple form
\be
{\cal J}^a = (j^a_{\bar z} - j^a_{z})\,.
\ee
When we choose a real form of the supergroup that has a compact subgroup,
the level $k$ will be integer.
We also observe that the current associated to the $\sigma$-component
of the canonical right-invariant one-form $dg g^{-1}$ is:
\be
{\cal J}'^a = c_- j_z + c_+ j_{\bar{z}}.
\ee
In term of these currents, we find the mode algebra%
\footnote{We thank A. Babichenko for pointing out a wrong sign in an earlier version.}%
:
\begin{eqnarray}
[{\cal J}^a_{n},{\cal J}^b_{m}]&=& -\frac{ c_1-c_3}{(c_2+c_4-g)^2} \kappa^{ab} n
  \delta_{m+n,0}-i {f^{ab}}_c {\cal J}^c_{n+m} \nonumber \\
{[} {\cal J}'^a_{n},{\cal J}^b_{m} {]}&=& -i \frac{c_1 c_- + c_3 c_+}{c_2+c_4-g}
\kappa^{ab} n   \delta_{m+n,0}-i{f^{ab}}_c {\cal J}'^c_{n+m}
\nonumber \\
{[} {\cal J}'^a_{n},{\cal J}'^b_{m} {]}&=& (c_-^2 c_1-c_+^2 c_3)
\kappa^{ab} n   \delta_{m+n,0}
+  {f^{ab}}_c 
(2 c_2 c_--2c_4 c_+ + g (c_+ - c_-))
{\cal J}'^c_{n+m}\nonumber \\
& & 
- i {f^{ab}}_c (c_2+c_4-g) (c_-^2 c_2 + c_4 c_+^2 - g(c_+^2 + c_+ c_- + c_-^2)) 
{\cal J}^c_{n+m}.
\end{eqnarray}
For the specific case of the supergroup model, we find 
\begin{align}
[{\cal J}^a_{n},{\cal J}^b_{m}]&= k \kappa^{ab} n
  \delta_{m+n,0}-i {f^{ab}}_c {\cal J}^c_{n+m} \cr
{[} {\cal J}'^a_{n},{\cal J}^b_{m} {]}&= -\frac{(k f^2-1)(k f^2+1)}{4 f^4}
\kappa^{ab} n   \delta_{m+n,0}-i{f^{ab}}_c {\cal J}'^c_{n+m}
\cr
{[} {\cal J}'^a_{n},{\cal J}'^b_{m} {]}&= 0 \,.
\end{align}
We identified a Kac-Moody subalgebra ${\cal J}$ and an infinite set of
 modes ${\cal J}'$ that commute amongst themselves. The latter modes transform into the identity
and themselves under the Kac-Moody algebra.

We also note that we can obtain a second Virasoro algebra by applying
the Sugawara construction to the Kac-Moody algebra ${\cal J}$. The
corresponding energy-momentum tensor generates a Virasoro algebra at
central charge $sdim\ G$. It is not holomorphic. The difference of
these energy momentum tensors for the left and right group is
proportional to
the difference of the holomorphic and
anti-holomorphic energy momentum tensors. That 
indicates
the existence of a non-chiral
analogue of the Knizhnik-Zamolodchikov equation.

\section{Conclusions}
\label{conclusions}
In this paper, we have performed a generic analysis of the conditions
imposed on local Lorentz covariant and $PT$ invariant current
algebras. In particular we allowed for parity-breaking models and
found a class of solutions to the conditions.

In the case for which the algebra has vanishing Killing form, we
showed that one can construct an energy momentum tensor in terms of a
current component in a way similar to the Sugawara construction.  The current
component is then a conformal primary and the central charge is the
(super)dimension of the group. This gives a constructive proof of
conformality of the quantum model.

We moreover computed exact two- and three-point functions for
principal chiral models with Wess-Zumino term for supergroups with
vanishing Killing form. Using these exact results, we 
showed that the current algebra is realized in these models, and we
calculated the coefficients in the current algebra. We performed a
check on a logarithmic regular term by computing the relevant part of
a four-point function. The algebra was independently derived by
using the techniques of conformal perturbation theory about
the Wess-Zumino-Witten point. We hope the existence of such current
algebras will prove useful in furthering the solution of these models
\cite{Read:2001pz, Gotz:2006qp, Quella:2007sg}. Another avenue
to explore is to systematically analyze the exactness of low-order
perturbation theory for various current and group valued
correlators.

One of the examples to which our discussion applies is the sigma model
on the supergroup $PSU(1,1|2)$.
This particular supergroup is useful to quantize string
theory on $AdS_3 \times S^3$ \cite{Berkovits:1999im, Bershadsky:1999hk}. To quantize the
string in the presence of Ramond-Ramond fluxes, we can, in this instance, use the
six-dimensional hybrid formalism with eight \cite{Berkovits:1994vy} or
sixteen \cite{Berkovits:1999du} manifest supercharges. In the first case, the
$PSU(1,1|2)$ sigma-model is at the core of the worldsheet theory
\cite{Berkovits:1999im}.

It is possible to realize the $AdS_3 \times S^3$ spacetime as the
near-horizon limit of a D5-NS5-D1-F1 system. We can then write the
parameters of the non-chiral current algebra in terms of the numbers
of D5 and NS5 branes \cite{Berkovits:1999im}. The integer parameter
$k$ that multiplies the Wess-Zumino term in the action is equal to the
number $N_{NS5}$ of NS5 branes while the parameter $1/f$ is the radius of
curvature of spacetime. When the number $N_{D5}$ of D5-branes is equal to zero, the parameters satisfy $k f^2=1$ and the non-holomorphic component of the right-invariant current
vanishes : we have a chiral current algebra. When we turn on the RR
fluxes, we obtain the generic current algebra given in 
equation \eqref{defalgebralogs}.
It is important to further investigate this algebra 
 in the context of string theory on $AdS_3$. Exploring
the integrability of these supergroup models will prove
useful in understanding better the properties of the $AdS_3/CFT_2$
correspondence.
The presence of a Kac-Moody algebra at level $k$ over the whole moduli
space of the theory may also help in  the construction of the string
spectrum in $AdS_3 \times S^3$ with Ramond-Ramond fluxes.

Likewise, another application of our analysis is to coset models $G/H$
where $G$ is a supergroup with zero Killing form.  In
\cite{Babichenko:2006uc} it was shown that a number of coset models
where $H$ is a maximal regular subalgebra are conformal to two loops.
Graded supercosets based on supergroups with vanishing Killing form
are also believed to be conformal \cite{Kagan:2005wt}. These
cosets occur in the worldsheet description of certain string theory
backgrounds, for instance, they appear as the central building block
of the $AdS_5 \times S^5$ background. Moreover, as symmetric spaces or
right coset manifolds, they retain a left group action as a symmetry
and we therefore expect that parts of our analysis still apply.
It is certainly worth exploring the quantum integrability of these coset
models per se, and how it ties in with the conformal current algebra
that we have exhibited.

\section*{Acknowledgements}
We would like to thank Costas Bachas, Zaara Benbadis, Denis Bernard,
Christian Hagendorf, Christoph Keller, Andre LeClair, Giuseppe
Policastro, Thomas Quella and Walter Troost for discussions.
We are grateful to Matthias Gaberdiel, Anatoly Konechny, Thomas Quella and an
anonymous referee for comments and corrections.

\begin{appendix}

\section{Perturbed operator product expansions}
\label{perturbOPE}

We consider the corrections to an OPE induced by an exactly marginal deformation of a 
conformal field theory.
The deformation parameter is denoted by $\lambda$.
In the deformed theory, we can
write the OPE between two operators $A$ and $B$ as: \be \lim_{z\to w} A(z) B(w) =
C(z,w) = \sum_{n\ge 0} \lambda^n C_n(z-w,w)\,, \ee where it is implicit
that the dependence on the variables does not have to be
holomorphic. We expand the result in a basis of operators
evaluated at the point $w$. The operator $C_n(z-w,w)$ is usually
written as a series in powers of $z-w$. It is not obvious that the
operators $A$ and $B$ (and therefore $C$) are well-defined operators
in the perturbed conformal field theory, but we will assume 
that that is the case for the model at hand.  
Let us see how to compute the operators
$C_n(z-w,w)$ at order $n$. By definition, we have: \be \lim_{z\to w}
\langle A(z) B(w) \phi_1(x_1) ... \phi_p(x_p) \rangle_\lambda =
\langle \left( \sum_{n\ge 0} \lambda^n C_n(z-w,w) \right) \phi_1(y_1)
... \phi_p(y_p) \rangle_\lambda \ee for any operators $\phi_1(y_1) ...
\phi_p(y_p)$.  If we want to perform the computation at the
non-perturbed point, we write the previous equality as: 
\begin{multline} 
\lim_{z\to w} \langle A(z) B(w) \phi_1(y_1) ...  \phi_p(y_p)
\left( \sum_{m\ge 0} \frac{\lambda^m}{m!} \prod_{i=1}^m \int d^2 x_i \Phi(x_i) \right) \rangle_0 \cr
= \langle \left( \sum_{n\ge 0} \lambda^n C_n(z-w,w) \right)
\phi_1(y_1) ... \phi_p(y_p) \left( \sum_{m\ge 0} \frac{\lambda^m}{m!} \prod_{i=1}^m \int d^2 x_i \Phi(x_i) \right)\rangle_0 
\end{multline}
where $\Phi$ is the exactly marginal operator we use to deform the theory.
We isolate the term proportional to $\lambda^n$: 
\begin{multline} 
\lim_{z\to w} \langle A(z) B(w) \phi_1(y_1) ... \phi_p(y_p)
\frac{1}{n!} \prod_{i=1}^n \int d^2 x_i \Phi(x_i)  \rangle_0 \cr
= \langle \left( \sum_{l=0}^n C_l(z-w,w) \frac{1}{(n-l)!}
  \prod_{i=1}^{n-l} \int d^2 x_i \Phi(x_i) \right) \phi_1(y_1) ...
\phi_p(y_p)\rangle_0 
\end{multline}
This becomes an operator identity in the
non-perturbed theory: \be \lim_{z\to w} A(z) B(w) \frac{1}{n!}
\prod_{i=1}^n \int d^2 x_i \Phi(x_i) = \sum_{l=0}^n C_l(z-w,w)
\frac{1}{(n-l)!} \prod_{i=1}^{n-l} \int d^2 x_i \Phi(x_i)\,. \ee The
previous equation defines iteratively the operator $C_n$ which appears
in the operator product expansion at order $n$.

We would like to give a prescription to compute the $n$-th order term in the OPE, $C_n(z-w,w)$. At zeroth order, the definition is
\be  
\lim_{z\to w}  A(z) B(w)   =   C_0(z-w,w)  \,.
\ee
As expected the zeroth-order OPE is the OPE in the non-deformed model.
At order one, we have \be \lim_{z\to w} A(z) B(w) \int d^2 x \Phi(x) =
C_0(z-w,w) \int d^2 x \Phi(x) + C_1(z-w,w) \ee Here is one proposal on how
to deal with the left-hand side of this equation. First we let the
operator $A(z)$ approach $B(w)$ and $\Phi(x)$ (separately): \be
\lim_{z\to w} A(z) B(w) \int d^2 x \Phi(x) = \lim_{z\to w} \left(
  (AB)(z-w,w) \int d^2 x\Phi + B(w)\int d^2 x (A\Phi)(z-x,x) \right)
\ee
where $(AB)(z-w,w)$ denotes the contraction of $A(z)$ and $B(w)$
(in the unperturbed theory), with the resulting operators evaluated at the point
$w$. It is clear that the first term on the right-hand side is equal to
$ C_0(z-w,w) \int d^2 x \Phi(x)$, so the OPE at first-order is given by
the second term: 
\be 
C_1(z-w,w) = \lim_{z\to w} B(w)\int d^2 x (A\Phi)(z-x,x) \,.
\ee 
At higher order, the same structure appears. We can always recognize in the computation the lower-order contributions, and isolate the highest-order term. We use the definition
\be 
\lim_{z\to w} A(z) B(w) \frac{1}{n!} \prod_{i=1}^n \int d^2 x_i \Phi(x_i) = \sum_{l=0}^n
C_l(z-w,w) \frac{1}{(n-l)!} \prod_{i=1}^{n-l} \int d^2 x_i \Phi(x_i) \,.
\ee
To evaluate the left-hand side, we let the operator $A(z)$ approach
the other ones. As it approaches $B(w)$, we generate the
term with $l=0$ on the right-hand side. As it approaches one of
the copies of the marginal operator $\Phi$, we get
\be \lim_{z\to w} B(w) \frac{1}{n!} \int d^2x (A\Phi)(z-x,x) \prod_{i=1}^{n-1} \int d^2 x_i \Phi(x_i) 
\ee 
To carry on, we take the operators $(A\Phi)(z-x,x)$ that was just
generated at the point $x$ and let it approach the other operators
in the expression. If it approaches $B(w)$, then we generate
the term with $l=1$ in the right-hand side of the definition. Otherwise we
generate a new expression on which we apply the same procedure.

Finally, we understand how to obtain directly the order-$n$ OPE
$C_n(z-w)$: it is the term that we get by first contracting $A(z)$
with all the integrated operators, and then contracting with $B(w)$ at the very
end. We will denote it as:
\be C_n(z-w,w) = \left[  A(z) \frac{1}{n!} \prod_{i=1}^n \int d^2 x_i \Phi(x_i) \right] B(w)\ee
All the operators inside the square brackets have to be contracted,
before performing the last contraction with the operator outside the
square brackets.

We should stress that the previous procedure is not always
well-defined. In the computation described in the bulk of this paper,
this prescription leads to an unambiguous result for the poles of the
current-current OPEs. However, in a more general context, the integrals appearing in the above
calculations need a more careful regularization.

\section{Detailed operator product expansions}\label{WZWOPE}

In this appendix we show how to compute OPEs involving the operator
$(g \bar J g^{-1})$ in the WZW model.

\subsubsection*{The contact terms}
It is natural to postulate contact terms between the left- and
right-invariant currents (see e.g. \cite{Kutasov:1988xb}).  Indeed, even for
a $U(1)$ current algebra, contact terms can be derived from the
representation of the current algebra in terms of a free boson and its
logarithmic propagator.  Since at large level $k$, the group manifold
flattens and is equivalent to a set of free fields,  we do
expect contact terms to arise. We propose the following contact terms:
\be 
J^a(z,\bar z) (g \bar{J} g^{-1})^b(w,\bar w)  \sim 2 \pi k \kappa^{ab}  \delta^{(2)}(z-w)  + ...
\ee

\subsubsection*{The Maurer-Cartan equation in the quantum theory}

In the quantum theory, the composite operator in the Maurer-Cartan equation is
ambiguous due to normal ordering. With our choice of normal
ordering, it is natural to propose the quantum
Maurer-Cartan equation
\be\label{MC}
\bar{\partial} J^c + \partial (g \bar{J} g^{-1})^c 
+ \frac{i}{2k} {f^{c}}_{de}  (:J^d (g \bar{J} g^{-1})^e: + : (g \bar{J} g^{-1})^e J^d :) = 0 \,.
\ee
One way to check this proposal is to compute the OPE between the current components $J^a$ and the operator on the left hand side of equation (\ref{MC}) which is classically zero due to the Maurer-Cartan
equation. In the calculation, it is crucial to apply the normal ordering prescription we introduced in section \ref{generic}.  
We not only confirm the above proposal for the quantum Maurer-Cartan equation, but also find that we need to fix the OPE between $J^a$ and
$(g \bar{J} g^{-1})^b$ to be
\be 
J^a(z)(g \bar{J} g^{-1})^b(w,\bar w) \sim 2\pi \kappa^{ab} \delta^{(2)}(z-w) 
+i {f^{ab}}_c \frac{(g \bar{J} g^{-1})^c(w,\bar w)}{z-w} + :J^a(g \bar{J} g^{-1})^b:(w,\bar w)\,.
\ee
Let us show this calculation in some detail in order to illustrate
 the techniques involved. Using the holomorphy of the current $J$ and
the knowledge of the naive conformal dimensions of the operators,
we can make the ansatz
\be J^a(z) (g \bar{J} g^{-1})^b(w,\bar w) = 2\pi k \kappa^{ab}
\delta^{(2)}(z-w) + \alpha i f^{abc} \frac{(g \bar{J} g^{-1})^c(w,\bar
  w)}{z-w} + : J^a (g \bar{J} g^{-1})^b:(w,\bar w) \ee
With the definition of the normal ordering above, let us compute the operator product expansion between $J^a(z)$ and the Maurer-Cartan equation. We distinguish two terms. The first term is
\begin{multline}\label{LHSMCJ} 
J^a(z) \left( \bar \partial J^c(w) + \partial (g \bar{J} g^{-1})^c(w, \bar w) \right) = \cr
\bar \partial_w \left(\frac{k \kappa^{ac}}{(z-w)^2} + if^{acd}\frac{J^d(w)}{z-w} \right)
  + \partial_w \left( 2\pi k \kappa^{ac} \delta^{(2)}(z-w) + \alpha i
    f^{acd} \frac{(g \bar{J} g^{-1})^d(w,\bar w)}{z-w}\right) + \ldots \cr
= - if^{acd} J^d(w) 2\pi
  \delta^{(2)}(z-w) + \alpha if^{acd}\frac{(g \bar{J} g^{-1})^d(w,\bar
    w)}{(z-w)^2}\cr
+ i f^{acd} \frac{\big( \alpha\,
    \partial (g \bar{J} g^{-1})^d(w,\bar w) + \bar \partial
    J^d(w)\big)}{z-w} + \ldots
\end{multline}
From the last terms we see that we can obtain a pole term proportional to the Maurer-Cartan equation if we put $\alpha=1$. It can be shown that this is the only consistent  possibility, and we will freely put $\alpha=1$ from now on.
The second term with an extra minus sign is given by
\be\label{MCRHS}
 -\frac{if^c_{de}}{k}J^a(z) \lim_{:x\rightarrow w:} (J^d(x) (g \bar{J} g^{-1})^e(w,\bar w)  + (g \bar{J} g^{-1})^e(w,\bar w)J^d(x) )\,.
\ee
where we have used the normal ordering prescription. Let us start with the first of the two terms in \eqref{MCRHS} (suppressing the overall $-\frac{i}{k}f^c_{de}$):
\begin{multline}
J^a(z) \lim_{:x\rightarrow w:} J^d(x) (g \bar{J} g^{-1})^e(w,\bar w) 
  = \lim_{:x\rightarrow w:}\left\{ \left[\frac{k \kappa^{ad}}{(z-x)^2}+i\frac{f^{ad}_g
  J^g(x)}{(z-x)} \right] (g \bar{J} g^{-1})^e(w,\bar w) \right. \cr
  \left. + J^d(x)\left[2\pi k \kappa^{ae}\delta^{(2)}(z-w) +
      i\alpha\frac{f^{ae}_g (g \bar{J} g^{-1})^g(w,\bar w)}{(z-w)}
    \right]\right\} \,.
\end{multline}
We perform successive contractions, and subtract singular terms according to the normal
ordering procedure to obtain
\begin{multline}
-2i\pi f^{ca}_{d}\delta^{(2)}(z-w) J^d(w) -
  \frac{i(f^{ca}_d\delta^{d}_{h}-\frac{\alpha}{k} f^c_{de}f^{ad}_g
    f^{ge}_h)(g \bar{J} g^{-1})^h(w,\bar w)}{(z-w)^2} \cr 
 + \frac{f^c_{de}(f^{ad}_g: J^g (g \bar{J} g^{-1})^e:(w,\bar w)+\alpha
    f^{ae}_g: J^d (g \bar{J} g^{-1})^g:(w,\bar w))}{k(z-w)}\,.
\end{multline}
When $\alpha =1$ and using the Jacobi identity, we can simplify further:
 \be
-2\pi i f^{ca}_d \delta^{(2)}(z-w)J^d(w) -i\left(1+\frac{2\hat{h}}{k}\right)\frac{f^{ca}_d (g \bar{J} g^{-1})^d}{(z-w)^2}-\frac{f^{ca}_d f^{d}_{ge} : J^g (g\bar J g^{-1})^e:(w,\bar w)}{k(z-w)}\,. 
\ee
Analogously, the second part of the second term becomes
\begin{multline}
- 2\pi i f^{ca}_d \delta^2(z-w) J^d(w) -i\left(1-\frac{2\hat{h}}{k}\right)\frac{(f^{ca}_{d} (g \bar{J} g^{-1})^d}{(z-w)^2}\cr
  +\frac{f^{c}_{de}}{k(z-w)}(f^{ae}_{g} :(g \bar{J} g^{-1})^g J^d:(w,\bar w) + f^{ad}_g :(g \bar{J} g^{-1})^e J^g:(w,\bar
  w))    \,.
\end{multline}
Combining the two parts of the second term we find 
\begin{multline}
2\pi i f^{ac}_d \delta^2(z-w) J^d(w) + \frac{if^{ac}_{d}(g \bar{J} g^{-1})^d}{(z-w)^2} \cr+ \frac{f^{ac}_{d} f^{d}_{eg}\big(:(g \bar{J} g^{-1})^g J^e:(w,\bar w) + :J^e (g \bar{J} g^{-1})^g :(w,\bar w)\big) }{2k(z-w)} \,.
\end{multline}
Comparing with the first term, we see that the contact term as well as
the double pole term cancel exactly while the single pole term
vanishes using the Maurer-Cartan equation itself, normal ordered as in
our proposal. We note that the operator product expansion between
$J^a$ and $(g \bar{J} g^{-1})^b$ obtained this way matches the
operator product expansion obtained in \eqref{finiteJXOPE} at the
Wess-Zumino-Witten point.

We will also need the OPE of $(g \bar{J} g^{-1})^a(z,\zbar)$ with
itself at the Wess-Zumino-Witten point:
\be
(g \bar{J} g^{-1})^a(z,\zbar) (g \bar{J} g^{-1})^b(w,\wbar) \sim \frac{k\kappa^{ab}}{(\zbar-\wbar)^2} +\frac{if^{ab}_c(z-w) J^c(w)}{(\zbar-\wbar)^2} - 2\frac{if^{ab}_c (g \bar{J} g^{-1})^c(w,\wbar)}{\zbar-\wbar} \,. 
\ee
The coefficients can be argued for by analyzing the contact and most singular terms in the OPE of the current $g \bar{J} g^{-1}$ with the Maurer-Cartan equation.

\subsection*{Operator product expansions of currents with marginal operator}

Since the computations are fairly similar, let us consider the more
complicated OPE of the current $(g \bar{J} g^{-1})^a(w,\wbar)$ with
the marginal operator $\Phi$. The first part of the computation
involves the OPE
\begin{multline}
  (g \bar{J} g^{-1})^b(w,\wbar)\lim_{:y\rightarrow x:} J^c(y)(g
  \bar{J} g^{-1})_c(x,\xbar) \sim \cr
  \lim_{:y\rightarrow x} \left\{\left(2\pi k \delta^{(2)}(w-y)+\frac{if^{bc}_d(g \bar{J}
        g^{-1})^d(y,\bar y)}{w-y}\right)(g\bar J g^{-1})_c(x,\xbar) \right. \cr
  \left.
    +J^c(y)\left(\frac{k\delta^b_c}{(\wbar-\xbar)^2}+\frac{if^b_{cd}(w-x)J^d(x)}{(\wbar-\xbar)^2}
      - \frac{2i f^b_{cd}(g \bar{J}
        g^{-1})^d(x,\xbar)}{\wbar-\xbar}\right) \right\}\cr \sim 2\pi
  k\delta^{(2)}(w-x)(g \bar{J}
  g^{-1})^b(x,\xbar)+\frac{kJ^b(x)}{(\wbar-\xbar)^2}-\frac{2if^b_{cd}}{(\wbar-\xbar)}:J^c(g
  \bar{J} g^{-1})^d:(x,\xbar)\cr
  +\frac{if^b_{cd}(w-x)}{(\wbar-\xbar)^2}:J^cJ^d:(x)-\frac{if^b_{cd}}{w-x}:(g
  \bar{J} g^{-1})^c(g \bar{J} g^{-1})^d:(x,\xbar)\,.
\end{multline}
Similarly, exchanging the order of $J$ and $g \bar{J} g^{-1}$, we find
an identical OPE to the above one except that the terms in the last
line have opposite sign. Combining these two, we therefore find that
\begin{multline}
(g \bar{J} g^{-1})^b(w,\wbar)\Phi(x,\xbar) \sim 2\pi k\delta^{(2)}(w-x)(g \bar{J} g^{-1})^b(x,\xbar)+\frac{k J^b(x)}{(\wbar-\xbar)^2}\cr
-\frac{2if^b_{cd}}{\wbar-\xbar}(:(g \bar{J} g^{-1})^dJ^c:+:J^c(g \bar{J} g^{-1})^d:)(x,\xbar)\,.
\end{multline}
Now, the last term can be rewritten using the Maurer-Cartan identity and we obtain
\begin{multline}
(g \bar{J} g^{-1})^b(w,\wbar)\Phi(x,\xbar) \sim 2\pi k\delta^{(2)}(w-x)(g \bar{J} g^{-1})^b(x,\xbar)+\frac{k J^b(x)}{(\wbar-\xbar)^2}\cr+\frac{2k}{\wbar-\xbar}(\bar\p J^b+\p (g \bar{J} g^{-1})^b)(x,\xbar)\,.
\end{multline}
Integrating over the location of the marginal operator and using the identities
\begin{align}
\int d^2 x \frac{\bar\p J^b}{\wbar-\xbar} &= -\int d^2x \frac{J^b(x)}{(\wbar-\xbar)^2} \\
\int d^2x \frac{\p (g \bar{J} g^{-1})^b}{\wbar-\xbar} &= 2\pi (g \bar{J} g^{-1})^a(w,\wbar)\,,
\end{align}
we find the 
contraction
\be
(g \bar{J} g^{-1})^b(w,\wbar)\int d^2x \Phi(x,\xbar) = 6\pi k (g \bar{J} g^{-1})^b(w,\wbar)-k\int d^2x \frac{J^b(x)}{(\wbar-\xbar)^2} \,.
\ee
For the OPE of $J^a(w)$ with the marginal operator, it turns out that
both orderings lead to the same answer, so we only exhibit the
following OPE:
\begin{align}
J^b(w)\lim_{:y\rightarrow x:} :J^c(y)(g \bar{J} g^{-1})_c(x,\xbar): &\sim \lim_{:y\rightarrow x:}\left\{ \left(\frac{k \kappa^{bc}}{(w-y)^2}+\frac{i f^{bc}_d J^d(y)}{w-y}\right)
(g \bar{J} g^{-1})_c(x,\xbar)\right. \cr
&\hspace{1cm}\left. +J^c(y)\left(2\pi k \delta^{(2)}(w-x)\delta^b_c + \frac{if^b_{cd}(g \bar{J} g^{-1})^d(x,\xbar)}{w-y} \right)\right\} \cr
&\sim \frac{k(g \bar{J} g^{-1})^b(x,\xbar)}{(w-y)^2}+2\pi k \delta^{(2)}(w-x)J^b(x)\cr
&\hspace{1.5cm}+\frac{if^b_{cd}}{w-y}(:J^d(g \bar{J} g^{-1})^c:+:J^c(g \bar{J} g^{-1})^d:)(x,\xbar)\cr
&\sim 2\pi k \delta^{(2)}(w-x)J^b(x) +\frac{k(g \bar{J} g^{-1})^b(x,\xbar)}{(w-y)^2}.
\end{align}

\section{Useful integrals}

We tabulate a few useful integrals that have been used throughout the article
(see e.g. \cite{Kutasov:1988xb}):
\begin{align}
\int  \frac{d^2 x}{(\xbar-\wbar)(x-z)}&= -2\pi \log|z-w|^2 \label{intone}\\
\int \frac{d^2x}{(\bar x - \bar w)^2(x-z)} &=2\pi \frac{1}{\bar z-\bar w} \label{inttwo}\\
\int \frac{d^2x}{(\bar x - \bar w)(x-z)^2} &= -2\pi \frac{1}{z-w} \label{intthree}\\
\int \frac{d^2x}{(\bar x - \bar w)^2(x-z)^2} &=4 \pi^2 \delta^{(2)}(z-w)\label{intfour} \\
\int \frac{ d^2 x}{(z-x)^2(w-x)} &= -2\pi \frac{\bar z - \bar w}{(z-w)^2} \label{intfive}
 \,.
\end{align}

\end{appendix}

\end{document}